\documentclass[10pt]{article}
\usepackage[utf8]{inputenc}
\usepackage{amsthm,amsmath,amssymb,amsfonts}
\usepackage{longtable}
\usepackage{hyperref}
\usepackage{times} % for times nr
\usepackage{graphicx}
\usepackage{geometry}
\usepackage{afterpage}%help with landscape layout
\usepackage{everypage}%add page number for landscape pages at the bottom
\usepackage{pdfpages}
\usepackage[affil-it]{authblk}


\title{\textbf{Global estimation and scenario-based projections of sex ratio at birth and missing female births using a Bayesian hierarchical time series mixture model}}
\date{\today} 

\author{Fengqing Chao\thanks{Corresponding author; Email: \href{mailto:fengqing.chao@kaust.edu.sa}{\texttt{fengqing.chao@kaust.edu.sa}}\\
This work is supported by a research grant from the National University of Singapore. The study described is solely the responsibility of the authors and does not necessarily represent the official views of the United Nations.\\
License: CC BY-NC-SA 4.0}}
\affil{Computer, Electrical and Mathematical Sciences and Engineering Division, King Abdullah University of Science and Technology (KAUST), Thuwal, Saudi Arabia\vspace{1em}}

\author{Patrick Gerland}
\affil{Population Estimates and Projections Section, United Nations Population Division, Department of Economic and Social Affairs, United Nations, New York, NY, USA\vspace{1em}}

\author{Alex R. Cook}
\affil{Saw Swee Hock School of Public Health, National University of Singapore and National University Health System, Singapore\vspace{1em}}

\author{Leontine Alkema}
\affil{Department of Biostatistics and Epidemiology, School of Public Health and Health Sciences, University of Massachusetts, Amherst, MA, USA\vspace{1em}}

\begin{document}
\pagenumbering{gobble}
\maketitle

\begin{center}
\textbf{Abstract}
\end{center}

The sex ratio at birth (SRB) is defined as the ratio of male to female live births. The SRB imbalance in parts of the world over the past several decades is a direct consequence of sex-selective abortion, driven by the co-existence of son preference, readily available technology of prenatal sex determination, and fertility decline. Estimation and projection of the degree of SRB imbalance is complicated because of variability in SRB reference levels and because of the uncertainty associated with SRB observations.

We develop Bayesian hierarchical time series mixture models for SRB estimation and scenario-based projections for all countries from 1950 to 2100. We model the SRB regional and national reference levels, and the fluctuation around national reference levels. We identify countries at risk of SRB imbalances and model both (i) the absence or presence of  sex ratio transitions in such countries and, if present, (ii) the transition process. The transition model of SRB imbalance captures three stages (increase, stagnation and convergence back to SRB baselines). The model identifies countries with statistical evidence of SRB inflation in a fully Bayesian approach. The scenario-based SRB projections are based on the sex ratio transition model with varying assumptions regarding the occurrence of a sex ratio transition in at-risk countries. Projections are used to quantify the future burden of missing female births due to sex-selective abortions under different scenarios. 

\clearpage

\pagenumbering{roman}
\tableofcontents

%\begin{spacing}{0.96}
\clearpage
%\listoftables
%\listoffigures 

\pagenumbering{arabic}
\setcounter{page}{1}

\section{Introduction}
Under normal circumstances, the sex ratio at birth (SRB; defined as the ratio of male to female live births) falls within a narrow range of 1.03 to 1.07 and varies slightly by ethnicity 
\cite{chahnazarian1988determinants,
chao2019systematic,
dubuc2007increase,
garenne2002sex,
garenne2008poisson,
graffelman2000statistical,
james1984sex,
james1985sex,
james1987human,
kaba2008sex,
ruder1985paternal,
marcus1998changing,
mathews2005trend,
visaria1967sex}. For most of human history, SRB remained within that natural range. However, over recent decades, SRBs have risen in a number of Asian countries and in Eastern Europe 
\cite{basten2013maternity,
bongaarts2013implementation,
bongaarts2015many,
chao2019systematic,
chao2019levels,
chen2020quantity,
choi2020transition,
das2003preference,
duthe2012high,
goodkind2011child,
guilmoto2007watering,
guilmoto2009recent,
guilmoto2009sex,
guilmoto2011socio,
guilmoto2012sex,
guilmoto2012skewed,
guilmoto2012preference,
hudson2004bare,
lin2009decline,
mesle2007sharp,
park1995consequences,
tafuro2020skewed,
vu2020sex}. 
SRB imbalance results from the interaction of three main factors \cite{guilmoto2009sex,guilmoto2012sex}: first, prolonged strong son preference, offering the motivation; second, fertility decreases leading to fewer children per family, inducing the willingness; third, accessible affordable sex-selection technology, providing the means. As a result, couples seek abortion based on the knowledge of the sex of the pregnancy to obtain sons while maintaining a small family size.

Estimation of the degree of SRB imbalance is challenging because of (i) variation in baseline SRB levels and (ii) uncertainty associated with SRB observations. In prior work \cite{chao2019systematic}, we developed a model to estimate SRB and imbalances for 212 countries from 1950 to 2017. The SRB estimation model accounts for the difference in the SRB reference levels across regions and varying uncertainty associated with SRB observations. We identified 29 countries where SRB imbalance may have happened in the past or may happen in the future, which we refer to as countries at risk of SRB inflation. We fitted a model for SRB levels and trends in country-years without risk of sex-selective abortion and obtained estimates for national and regional SRB baseline values. Subsequently, we estimated sex imbalances using a sex ratio transition model to capture periods of increasing, constant and decreasing sex ratio imbalances.

Constructing SRB projections is challenging for countries with ongoing SRB imbalances \cite{gupta2009evidence} and even more so for countries with normal SRB levels and trends but with potential of rising SRB in the future \cite{bongaarts2015many}. While prior work has assessed the potential future SRB imbalances for selected countries with ongoing transitions \cite{bongaarts2015many,chao2020probabilistic}, no work to date has quantified the possible additional SRB imbalances for countries where sex ratio transitions may start in the future. Efforts to date to project the SRB globally include those by the United Nations (UN) Population Division \cite{UNWPP2019}. The UN Population Division publishes projections of demographic indicators including the SRB for all countries in the World Population Prospects (WPP) every 2 years. While the UN WPP projections of fertility, mortality and populations are probabilistic, SRB projections are deterministic and based on expert-based opinions \cite{UNWPP2019}. Specifically, WPP methods are based on the assumption that future SRB outcomes either remain at the same level as most recently observed or converge to 1.05 within the next 10--40 years, and future SRB imbalance is not assessed. Similarly, the Global Burden of Disease produces SRB estimates and projections for all countries but does not assess SRB imbalances and assumes that the SRB remains constant in projections from 2017 up to 2100 \cite{vollset2020fertility}.

In this study, we extend the SRB estimation model \cite{chao2019systematic} to produce scenario-based SRB projections, and associated SRB imbalances, till 2100 for all countries. The update in the SRB inflation model allows for identification of countries with past and ongoing SRB inflation in a fully Bayesian approach. We use the updated model to estimate the sex ratio transition for country-years with risk of SRB inflation. Subsequently, we construct scenario-based SRB projections that are based on the sex ratio transition model with varying assumptions regarding the occurrence of a sex ratio transition in at-risk countries.

This paper is organized as follows: we first summarize the data used for estimating the SRB in Section~\ref{sec_data}. Section~\ref{sec_methods} introduces the models for estimating the SRB, followed by the approaches used to produce scenario-based SRB projections. In Section~\ref{sec_results}, we present results regarding SRB baselines, past SRB imbalances, and scenario-based projections, as well as validation results. Finally, in Section~\ref{sec_discussion}, we end with a discussion of implications of our model and results, limitations, and possible future research.

\section{Data}\label{sec_data}
We produce SRB estimates for 212 countries\footnote{We use the term ``country'' to refer to populations that are considered as countries or areas in the UN classification.} with total population size greater than 90,000 as of 2017. An overview of observations by data source type is in Table~\ref{tab-datadist}. There are 10,835 data points available from 202 countries in total.

\begin{table}[htbp]
\caption[SRB observations by source type]{\textbf{SRB observations by source type.} DHS: Demographic and Health Surveys. Other DHS refer to non-standard DHS, including Special, Interim and National DHS, Malaria Indicator Surveys, AIDS Indicator Surveys, World Fertility Surveys, Reproductive Health Survey, Multiple Indicator Cluster Surveys, Pan Arab Project for Child Development and Pan Arab Project for Family Health. CRVS: civil registration vital statistics. SRS: sampling registration system.}
\label{tab-datadist}
\centering
\begin{tabular}{ c r r} \hline
\bf{Data Source Type} & \bf{\# Observations} & \bf{\# Country-Years} \\ \hline
Census & 48 & 48\\ 
DHS & 2,257 & 5,413\\ 
Other DHS & 1,392 & 3,662\\ 
Other & 142 & 222\\ 
CRVS/SRS & 6,996 & 7,257\\ \hline
\textbf{Total} & \textbf{10,835} & \textbf{16,602}\\ \hline
\end{tabular}
\end{table}

Data availability is summarized by source type in Table~\ref{tab-datadist}. We compile civil registration vital statistics (CRVS) data from the UN Demographic Yearbook and the Human Mortality Database, and sampling registration system (SRS) data for India, Pakistan and Bangladesh from annual reports. CRVS and SRS typically provide data on an annual basis, based on government administrative record from birth certificates. 
International survey data (Demographic and Health Surveys, World Fertility Surveys, Reproductive Health Survey, Multiple Indicator Cluster Surveys, Pan Arab Project for Child Development and Pan Arab Project for Family Health) were compiled from microdata when possible, and obtained from reports otherwise. Census and national-level survey data were obtained from reports. Censuses usually provide information on SRB for a retrospective period of 12 or 24 months prior to the survey date. Surveys collect data on recent births or full birth histories from women of reproductive ages for longer retrospective periods of 5 to 20 years before the survey date. Details on  data series by country and preprocessing are given in \cite{chao2019systematic,chao2019web}.

\section{Methods}\label{sec_methods}

\paragraph*{Notation}
We use lowercase Greek letters for unknown parameters and uppercase Greek letters for variables which are functions of unknown parameters. Where relevant for communicating model assumptions, we add hyper parameters associated with parameters in parentheses. I.e. for some model parameter $\phi$, the notation $\phi(\zeta)$ implies that the distribution for $\phi$ is parametrized using hyper parameter $\zeta$. We use super script $(g)$ to the left of a parameter and a model/scenario name to the right to denote posterior samples of that parameter from a specific model fit, i.e. $^{(g)}{\phi}^{\text{\tiny [M1]}}$ refers to the $g$-th posterior sample of $\phi$ as obtained from model fit M1. $\hat{\phi}^{\text{\tiny [M1]}}$ refers to the posterior median of $\phi$ based on model M1. Roman letters indicate variables that are known or fixed, including data (in lowercase) and estimates provided by other sources or the literature (in uppercase).

$\Theta_{c,t}$ denotes the main outcome of interest to be modeled, which is the SRB for country $c$ in year $t$. Observations are combined across countries over time and indexed by $i\in\{1, \cdots, n\}$; $c[i]$ refers to the country the $i$-th observation belongs to, $t[i]$ to the calendar year of the observation, and $s[i]$ to the data source type (Table~\ref{tab-datadist}) of the observation. $r[c]$ refers to the region that country $c$ belongs to.

\subsection{SRB model summary}\label{sec_srb_model}
We assume the $i$-th observed SRB $y_i$ follows a normal distribution on the log-scale: 
\begin{eqnarray}
\log(y_i)|\Theta_{c[i], t[i]},\omega_{s[i]} &\sim& \mathcal{N}\left( \log(\Theta_{c[i], t[i]}), \omega_{s[i]}^2 + v_i^2\right), \label{eq_datamodel}
\end{eqnarray}
The variance for the $i$-th log-scaled SRB observation $\log(y_i)$ is the sum of known stochastic/sampling variance $v_i^2$ and unknown non-sampling error variance $\omega_{s[i]}^2$ for data source type $s[i]$ (Section~\ref{sec_datamodel}). 

In its general form, the process model for SRB $\Theta_{c,t}$ is defined as follows:
\begin{eqnarray}
\Theta_{c,t} &=& \beta_c \eta_{c,t}(\boldsymbol{\phi}) + \delta_c \Omega_{c,t}(\boldsymbol{\zeta}), \label{eq_processmodel}
\end{eqnarray}
where $\beta_c$ refers to country-specific time-invariant baseline SRB, and $\eta_{c,t}$ the country-year-specific fluctuations around the baseline. $\delta_c$ indicates the absence or presence of SRB inflations in country $c$. For countries with inflations, $\Omega_{c,t}$ captures the SRB imbalance in year $t$ and can be interpreted as the additional number of male births for each female birth that is not aborted. The vector of hyper parameters related to $\eta_{c,t}$ is denoted as $\boldsymbol{\phi}$, and we use $\boldsymbol{\zeta}$ for the vector of hyper parameters related to $\Omega_{c,t}$.

We identify a set of countries where SRB inflation may have happened in the past or may happen in the future, i.e. countries with $\Pr(\delta_c >0) \neq 0$, which we refer to as ``countries at risk of SRB inflation'' (see Section~\ref{sec_method_step1}). In our study, we focus on sex ratio transitions that are due to sex-selective abortion, not other factors (e.g. natural disaster, economic crises, famine and war) that could result in acute changes in SRB \cite{song2012does,venero2011association,catalano2006exogenous,fukuda1998decline}. The estimation of baseline SRB $\beta_c$ and fluctuations around baseline $\eta_{c,t}$ is described in Section~\ref{sec_method_step2}. The sex transition model $\delta_c \Omega_{c,t}$ is given in Section~\ref{sec_method_step3}. Finally, in Section~\ref{sec_method_sceproj}, we introduce the approach for constructing scenario-based projections, using the models developed in the preceding sections. The full model specification,  priors and details on computation are in the Appendix~\ref{app_model}.

\subsection{Error variances}\label{sec_datamodel}
We account for differences in error variance across observations from CRVS, surveys and censuses. Errors---and hence the error variance---associated with non-CRVS data tend to be larger than errors associated with CRVS data and this is reflected in the model fitting, as the weight assigned to a data point increases as its error variance decreases. Resulting model-based estimates are more strongly weighted by observations with smaller errors, and uncertainty ranges are narrower for country-periods with more observations with smaller error variance. 

As per Equation~\ref{eq_datamodel}, the variance for the $i$-th log-scaled observed SRB $\log(y_i)$ is the sum of known stochastic/sampling variance $v_i^2$ and unknown non-sampling error variance $\omega_{s[i]}^2$ for data source type $s[i]$ as listed in Table~\ref{tab-datadist}. For CRVS observations, $v_i$ is the stochastic error and is pre-calculated as described elsewhere \cite{chao2019systematic,chao2019web}, and
we assume that non-sampling error is zero: $\omega_s=0$ when $s=\text{CRVS/SRS}$. For observations from surveys or censuses, $v_i$ is the sampling error and is pre-calculated using a jackknife method as explained in \cite{chao2019systematic,chao2019web}, to reflect the survey sampling design. Non-sampling variance term $\omega_s^2$ captures random errors that may occur during the data collection process. This variance parameter is estimated and assigned a vague prior:
\begin{eqnarray}
\omega_{s} &\sim& \mathcal{U}(0,0.5), \text{ for }s \in \{\text{Census, DHS, Other DHS, Other}\}.
\end{eqnarray}

\subsection{Selection of countries at risk of SRB inflation}\label{sec_method_step1}
The model for natural fluctuations in the SRB is fitted to the global database after excluding data from country-years that may have been affected by masculinisation of the SRB. We use inclusive criteria to identify such country-years, based on a combination of qualitative and quantitative approaches. We select countries with at least one of the following manifestations of son preference: (i) a high level of desired sex ratio at birth (DSRB), or (ii) a high level of sex ratio at last birth (SRLB), or (iii) strong son preference or inflated SRB suggested by a literature review \cite{chao2019systematic,chao2019web}. DSRB is calculated as the ratio of the reported number of desired male births to desired female births, as reported by women during survey interviews. The DSRB reflects the desired sex composition at the time of survey interview. The SRLB quantifies the sex ratio among births that are the latest births to women who desire no more children. 

A total of 29 countries satisfy at least one of the aforementioned criteria, and hence are considered at risk of SRB inflation (see \cite{chao2019systematic,chao2019web}): 
Afghanistan; Albania; Armenia; Azerbaijan; Bangladesh; China; Egypt; Gambia; Georgia; Hong Kong, SAR of China; India; Jordan; Republic of Korea; Mali; Mauritania; Montenegro; Morocco; Nepal; Nigeria; Pakistan; Senegal; Singapore; Taiwan, Province of China; Tajikistan; Tanzania; Tunisia; Turkey; Uganda; Vietnam.

\subsection{Model stage 1: estimating the SRB in country-years without SRB inflation}\label{sec_method_step2}
In the model for country-years not affected by sex-selective abortion, we assume $\delta_c=0$ in Equation~\ref{eq_processmodel}. The SRB is thus given by a product of two components:
\begin{eqnarray}
\Theta_{c,t} &=& \beta_c \eta_{c,t}(\boldsymbol{\phi}),\label{eq_basemodel}
\end{eqnarray}
where $\beta_c$ is a national baseline for country $c$, which is assumed to be constant over time and $\eta_{c,t}(\boldsymbol{\phi})$ is a country-year-specific multiplier that captures the natural fluctuation of the country-specific SRB around its respective baseline value over time, and $(\boldsymbol{\phi})$ is the vector of the hyper parameters related to $\eta_{c,t}$. We label above Equation~\ref{eq_basemodel} as ``M1'' to refer to model stage 1. In this step of modeling, parameters that are not related to prenatal gender discrimination and sex-selective abortion are estimated. To do so, we use a reduced SRB database, referred to as the risk-free database denoted by $\boldsymbol{y}^{(\text{risk-free})}$, by excluding SRB observations that may be affected by prenatal sex discrimination and sex-selective abortion. The risk-free database is obtained by excluding data points with reference years after 1970 from the 29 countries at risk of SRB inflation (listed in Section~\ref{sec_method_step1}) because the sex-selective abortion technology became more accessible and affordable since then \cite{allahbadia200250,george2002sex,goodkind1997sex,oomman2002sex,tandon2006female}. Data points with reference years before 1970 from the 29 countries and all data points from remaining countries are included in the risk-free database for fitting the model.

The national baseline $\beta_c$ follows a hierarchical distribution with mean at its corresponding regional baseline $\beta_{r[c]}^{(\text{region})}$:
\begin{eqnarray}
\log(\beta_c) | \beta_{r[c]}^{(\text{region})}, \sigma_\beta &\sim& \mathcal{N}\left(\log\left(\beta_{r[c]}^{(\text{region})}\right), \sigma_\beta^2\right).
\end{eqnarray}
%The hierarchical structure for national baseline $\beta_c$ is to account for ethnicity difference across countries within the same region $r[c]$. 
National baselines are pooled toward regional baseline $\beta_r^{(\text{region})}$ to capture SRB differences due to ethnic origin 
\cite{chahnazarian1988determinants,
dubuc2007increase,
	garenne2002sex,
	garenne2008poisson,
	graffelman2000statistical,
	james1984sex,
	james1985sex,
	james1987human,
	kaba2008sex,
	ruder1985paternal,
	marcus1998changing,
	mathews2005trend,
	visaria1967sex}. For example, we group countries in Europe, North America, Australia, and New Zealand to refer to the regional grouping of countries with a majority of Caucasians. We assume that the national baseline $\beta_c$ and regional baseline $\beta_r^{(\text{region})}$ are constant over time. We assign independent uniform priors to each $\beta_r^{(\text{region})}$ and a vague prior to $\sigma_\beta$.

The country-year-specific multiplier $\eta_{c,t}$ is modeled on the log-scale with an autoregressive AR(1) time series process within a country. For countries without any data or with very limited information, $\eta_{c,t}$ is shrunk towards 1, such that the estimated SRBs without prenatal sex discrimination are close to their corresponding national baselines $\beta_c$. For countries where the data suggest different levels or trends, $\eta_{c,t}$ captures these deviations from $\beta_c$. Let $\boldsymbol{\phi}=\{ \rho, \sigma_{\epsilon}\}$ be the vector of hyper parameters related to $\eta_{c,t}$. We assume:
\begin{eqnarray}
\log(\eta_{c,t}) | \boldsymbol{\phi} &\sim& \mathcal{N}(0, (1-\rho^2)/ \sigma_\epsilon^2), \text{ for }t=1950,\\
\log(\eta_{c,t}) &=& \rho\log(\eta_{c,t-1}) + \epsilon_{c,t}, \text{ for }t\in\{1951, \cdots, 2100\},\\
\epsilon_{c,t} | \sigma_\epsilon &\overset{\text{i.i.d.}}\sim& \mathcal{N}(0, \sigma_\epsilon^2).
\end{eqnarray}
Vague priors are assigned to $\sigma_\beta$, $\rho$ and $\sigma_\epsilon$.
%\begin{eqnarray*}
%\sigma_\beta &\sim& \mathcal{U}(0, 0.05).
%\rho &\sim& \mathcal{U}(0, 1),\\
%\sigma_\epsilon &\sim& \mathcal{U}(0, 0.05).
%\end{eqnarray*}

\subsection{Model stage 2: estimating the SRB in country-years at risk of SRB inflation}\label{sec_method_step3}
We model SRB $\Theta_{c,t}$ for the 29 countries at risk of SRB inflation (listed in Section~\ref{sec_method_step1}) as the sum of two parts: (i) the inflation-free SRB level, given by the model of country-years without SRB inflation as described in Section~\ref{sec_method_step2}; and (ii) the SRB imbalanced level with probability. We fit the model to all data from the 29 at-risk countries and let $\boldsymbol{y}^{(\text{at-risk})}$ denote the database in model stage 2. Specifically, $\Theta_{c,t}$ for country $c$, year $t$ is modeled as:
\begin{eqnarray}
\Theta_{c,t} &=& \hat{\beta}^{\text{\tiny [M1]}}_c \eta_{c,t}\left(\boldsymbol{\hat{\phi}}^{\text{\tiny [M1]}} \right) + \delta_c \Omega_{c,t}(\boldsymbol{\zeta}),\label{eq_inflatedsrb}
\end{eqnarray}
where $\hat{\beta}^{\text{\tiny [M1]}}_c$ is the posterior median estimate for the national baseline and $\boldsymbol{\hat{\phi}}^{\text{\tiny [M1]}}=\{ \hat{\rho}^{\text{\tiny [M1]}}, \hat{\sigma_{\epsilon}}^{\text{\tiny [M1]}} \}$ the vector of posterior medians of $\boldsymbol{\phi}$ obtained from the inflation-free model stage 1 (M1) fit described in Section~\ref{sec_method_step2}. We label above Equation~\ref{eq_inflatedsrb} as ``M2'' to refer to model stage 2. In M2, we impute point estimate $\hat{\beta}^{\text{\tiny [M1]}}_c$ from M1 rather than estimating $\beta_c$ and inflation $\Omega_{c,t}$ jointly to avoid upwards bias in the estimated start year of SRB inflation in China and India due to overestimation of $\beta_c$, see Appendix~\ref{app_model_jointseq}. 

The product $\delta_c \Omega_{c,t}(\boldsymbol{\zeta})$ captures SRB imbalance. It is the product of a binary indicator indicating presence or absence of SRB inflation $\delta_c$ for country $c$, and a non-negative SRB inflation $\Omega_{c,t}(\boldsymbol{\zeta})$ where $\boldsymbol{\zeta}$ is the vector of hyper parameters related to $\Omega_{c,t}$.

The country-specific binary factor $\delta_c$ detects the existence of SRB inflation, with values either 0 (no inflation) or 1 (with inflation). $\delta_c$ is modeled with a Bernoulli distribution with country-specific probability of having inflation $\pi_c$:
\begin{eqnarray}
\delta_c | \pi_c &\sim& \mathcal{B}(\pi_c).\label{eq_delta_bhm1}
\end{eqnarray}
Logit-transformed $\pi_c$ follows a hierarchical normal distribution with a global mean at $\mu_\pi$ and a global variance at $\sigma_\pi^2$: 
\begin{eqnarray}
\text{logit}(\pi_c) | \mu_\pi, \sigma_\pi &\sim& \mathcal{N}(\mu_\pi, \sigma_\pi^2).\label{eq_delta_bhm2}
\end{eqnarray}
Vague priors are assigned to $\sigma_\pi$ and $\mu_\pi$.

\begin{figure}
\begin{centering}
\includegraphics[width = 0.55\textwidth]{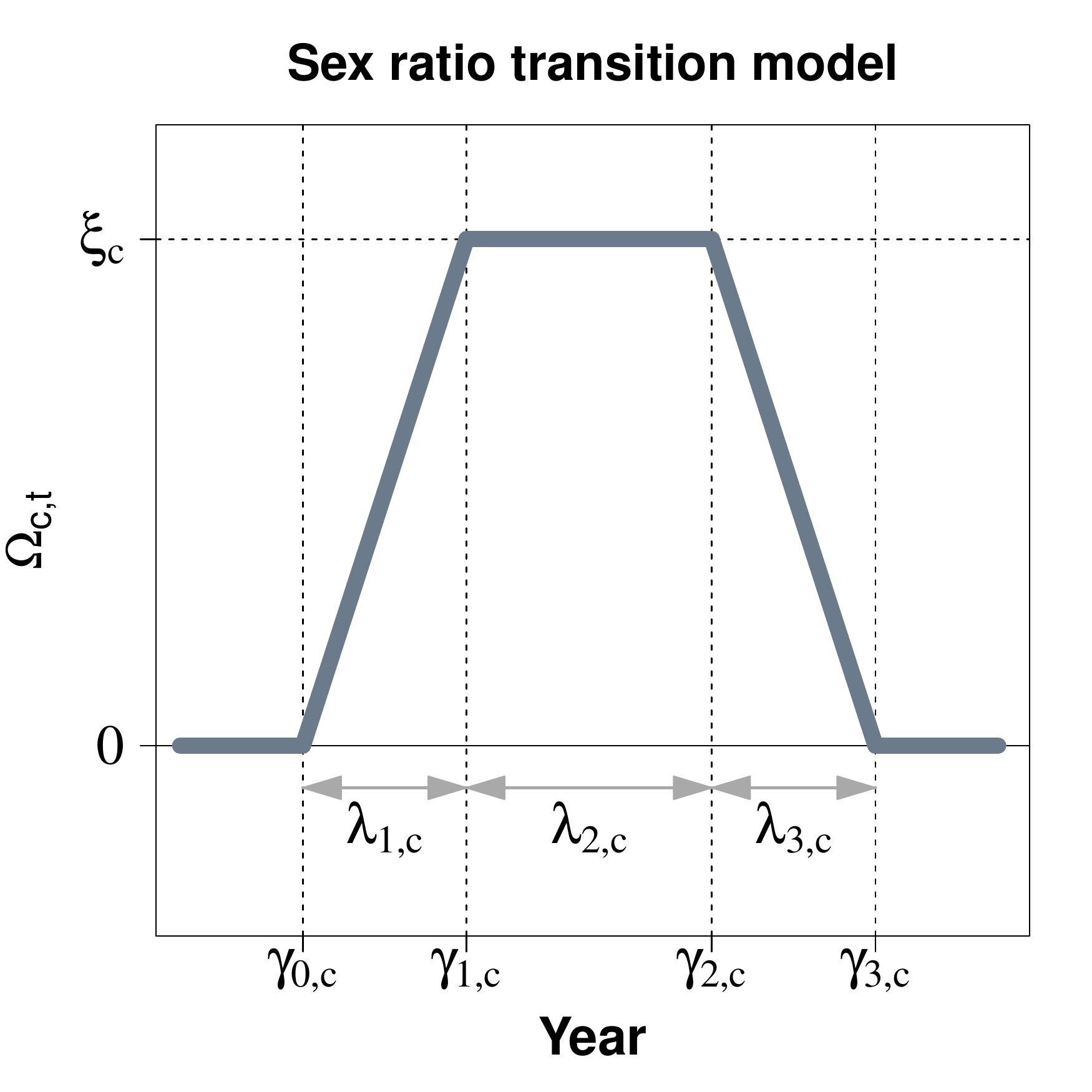}
\caption[Illustration of sex ratio transition model.]{\textbf{Illustration of sex ratio transition model.} All parameters are for country $c$. $\gamma_{0,c}$: start year of SRB inflation. $\lambda_{1,c}$, $\lambda_{2,c}$ and $\lambda_{3,c}$: period lengths of the SRB inflation phases of increase, stagnation and decrease back to zero. $\xi_c$: maximum value of the inflation.}\label{fig-adj-illustration}
\end{centering}
\end{figure}

$\Omega_{c,t}$ is the upward SRB inflation factor for country $c$ in year $t$ to capture higher SRB levels that may be due to sex-selective abortion. We parameterize the sex ratio transition using a trapezoid function to represent consecutive phases of increase, stagnation and decrease back to zero (Figure~\ref{fig-adj-illustration}). The inflation factor $\Omega_{c,t}$ is modeled as:
\begin{eqnarray}
\Omega_{c,t} &=&
\left\{\begin{matrix}
\xi_c (t - \gamma_{0,c}) / \lambda_{1,c}, & \gamma_{0,c} < t < \gamma_{1,c} \\ 
 \xi_c, & \gamma_{1,c} <t < \gamma_{2,c} \\
 \xi_c - \xi_c (t -\gamma_{2,c}) / \lambda_{3,c}, & \gamma_{2,c} < t < \gamma_{3,c} \\
0, & t < \gamma_{0,c} \text{ or } t > \gamma_{3,c}
\end{matrix}\right.,\text{ where}\\\nonumber \label{eq_Omega}
\gamma_{1,c} &=& \gamma_{0,c} + \lambda_{1,c},\\\nonumber
\gamma_{2,c} &=& \gamma_{1,c} + \lambda_{2,c},\\\nonumber
\gamma_{3,c} &=& \gamma_{2,c} + \lambda_{3,c},\nonumber
\end{eqnarray}
with sex ratio transition parameters $\gamma_{0,c}$, the start year of the inflation, $\xi_c$, the maximum inflation, and $\lambda_{1,c}$, $\lambda_{2,c}$ and $\lambda_{3,c}$, the lengths of the inflation period during the three phases. 

The model for $\Omega_{c,t}$ makes use of fertility as an external covariate related to SRB inflation to better capture the sex ratio transition process. Specifically, information related to the ``fertility squeeze'' effect (fertility decreases leading to fewer children per family, inducing the willingness to sex-selective abortion) is incorporated into the model through the parametrization of the start year $\gamma_{0,c}$ (see Section \ref{sec_inf_start}). Despite the acknowledged role of son preference (offering the motivation of sex-selective abortion) or availability of sex-selection technology (providing the means of sex selection) in driving SRB imbalance \cite{guilmoto2009sex,guilmoto2012sex}, the model for $\Omega_{c,t}$ does not make use of covariates related to these two factors because detailed information is general unavailable for estimation and available candidate predictors do not explain variability across countries \cite{chao2019systematic}. Instead, sex ratio transition parameters related to the lengths of the transition phases and its maximum are estimated with Bayesian hierarchical models \cite{gelman2013bayesian,lindley1972bayes}.
\begin{eqnarray}
\xi_c | \mu_{\xi}, \sigma_{\xi} &\sim& \mathcal{N}(\mu_{\xi}, \sigma_{\xi}^2)T(0, ), \label{eq_Omega_bhm1}\\
\lambda_{1,c} | \mu_{\lambda1}, \sigma_{\lambda1} &\sim& \mathcal{N}(\mu_{\lambda1}, \sigma_{\lambda1}^2)T(0, ), \label{eq_Omega_bhm2}\\
\lambda_{2,c} | \mu_{\lambda2}, \sigma_{\lambda2}&\sim& \mathcal{N}(\mu_{\lambda2}, \sigma_{\lambda2}^2)T(0, ), \label{eq_Omega_bhm3}\\
\lambda_{3,c} | \mu_{\lambda3}, \sigma_{\lambda3} &\sim& \mathcal{N}(\mu_{\lambda3}, \sigma_{\lambda3}^2)T(0, ), \label{eq_Omega_bhm4}
\end{eqnarray}
where $\mathcal{N}(\cdot)T(0, )$ refers to a truncated normal distribution with lower truncation at zero. The hierarchical models allow for projecting complete transitions in all countries, including countries that have not yet finished their transitions, based on (partially) observed transitions so far. We assign vague priors to the mean and standard deviation of these truncated distributions (Appendix~\ref{app_model}).

\subsubsection{Inflation start year}\label{sec_inf_start}
We model the SRB inflation start year $\gamma_{0,c}$ as:
\begin{eqnarray}
\gamma_{0,c}  | \sigma_{\gamma} &\sim& t_3(x_c, \sigma_{\gamma}^2)T(z_c, ),\label{eq_Omega_bhm5}
\end{eqnarray}
referring to a Student-$t_3$ distribution with lower truncation in year $z_c$, location $x_c$ and scale $\sigma_{\gamma}$. The fertility squeeze effect is incorporated into the model for $\gamma_{0,c}$ using country-year estimates of the total fertility rate $F_{c,t}$ from the UN World Population Prospects (WPP) 2019 \cite{UNWPP2019}. The total fertility rate (TFR) approximates the number of children that would be born to a woman who survives throughout her reproductive ages. The lower truncation $z_c$ refers to the year that the TFR in country $c$ decreased to 6 or the year 1970, whichever occurred later:
\begin{equation}
%    z_c = \max\{1970,\arg\min_{t \in \{1970,\cdots,2100\}}\{|\mathbf{F}_{c} - 6\mathbf{1}_{131}|\}\},
z_c = \max\{1970, f_{c,6}\},
\end{equation}
%where $\mathbf{1}_{131}$ is the vector with 131 entries and all entries are 1.
where $f_{c,6}$ is the first year in which the TFR in country $c$ declines to 6. The location indicator $x_c$ is the year in which the TFR in country $c$ declines to the pre-calculated value of 2.9 (the TFR value 2.9 is determined by the median of fertility levels when the observed sex ratio transition started among countries with high-quality CRVS data \cite{chao2019systematic,chao2019web}) or 1970, whichever occurred later: %$f_{c,2.9}$:
\begin{equation}
x_c = \max\{1970, f_{c,2.9}\}.
\end{equation}

\subsection{Scenario-based projections}\label{sec_method_sceproj}
We construct scenario-based SRB projections that are based on the sex ratio transition model in Equation~\ref{eq_processmodel} with varying assumptions regarding the occurrence of a sex ratio transition in at-risk countries. The three scenarios are:
\begin{itemize}
\item S1: SRB inflation continues in countries with strong statistical evidence of past inflations only.
\item S2: SRB inflations are projected to occur in all at-risk countries with country-specific inflation probabilities.
\item S3: SRB inflations are projected to occur in all at-risk countries with probability 1. 
\end{itemize}
The scenarios are labeled in terms of increasing sex imbalance in future years, with S1 including the sex imbalance due to ongoing transitions in countries with strong statistical evidence of such inflations only, and S3 including sex imbalance for at-risk countries without current evidence of inflation. For each country-year $c,t$, we let $^{(g)}{\Theta}^{\text{\tiny [S]}}_{c,t}$ denote the  $g$-th posterior sample of $\Theta_{c,t}$ for scenario S $\in \{\text{S1}, \text{S2}, \text{S3}\}$. 

We identify countries with strong statistical evidence of SRB inflation in a fully Bayesian approach, using $\psi_c$, the posterior mean of the inflation indicator $\delta_c$ for country $c$ from M2: $\psi_c = \mathbb{E}(\delta_c|\boldsymbol{y}^{(\text{at-risk})})$. 
$\psi_c$ represents the relative inclusion of sex ratio inflation in country $c$.
If $\psi_c \ge 95\%$, then we consider the country to have strong statistical evidence of SRB inflation. We use $\mathcal{C}^{\text{inflation}}$ to denote the set of country indices $c$ with $\psi_c \ge 95\%$.  

The projections are based on estimates from different model fits, summarized in Table~\ref{tab_modelfits}. For countries that are not at risk of SRB inflation, with $c \in \mathcal{C}^{\text{base}}$, projections are based on the country-specific baseline estimates and deviations away from that baseline and are the same across scenarios, obtained from M1. Specifically, $^{(g)}{\Theta}^{\text{\tiny [S]}}_{c,t}$, the $g$-th posterior sample of $\Theta_{c,t}$ for scenario S $\in \{\text{S1}, \text{S2}, \text{S3}\}$ is obtained as below:
\begin{equation}
^{(g)}{\Theta}^{\text{\tiny [S]}}_{c,t} = ^{(g)}{\beta}^{\text{\tiny [M1]}}_c \cdot ^{(g)}{\eta}^{\text{\tiny [M1]}}_{c,t},\text{ for scenario S}  \in \{\text{S1}, \text{S2}, \text{S3}\}, c \in \mathcal{C}^{\text{base}}.
\end{equation}
As indicated in the notation section, we introduce the  superscript $(g)$ on the left hand side and $[\text{M}]$ on the right hand side of a parameter to refer to posterior sample $g$ of that parameter from model $M$. 

\begin{table}[htbp]
\caption[Summary of models]{\textbf{Summary of models.} $\boldsymbol{y}^{(c)}$: all data points from country $c$. $\boldsymbol{\hat{\zeta}}^{\text{\tiny [M2]}}$ refers to point estimates of the vector of hyper parameters for the inflation factor from M2 ($\boldsymbol{{\zeta}} = \{\mu_{\xi}, \sigma_{\xi}, \mu_{\lambda1}, \sigma_{\lambda1}, \mu_{\lambda2}, \sigma_{\lambda2}, \mu_{\lambda3}, \sigma_{\lambda3}, \sigma_{\gamma}\}$).}
\label{tab_modelfits}
\centering
\begin{tabular}{|l|l|l|} \hline
\bf{Model} & \bf{Model equation} & \bf{Database}\\ \hline
M1 & $\Theta_{c,t} = \beta_c \eta_{c,t}(\boldsymbol{\phi})$ & $\boldsymbol{y}^{(\text{risk-free})}$ \\
M2 & $\Theta_{c,t} = \hat{\beta}^{\text{\tiny [M1]}}_c \eta_{c,t}\left(\boldsymbol{\hat{\phi}}^{\text{\tiny [M1]}} \right) + \delta_c \Omega_{c,t}(\boldsymbol{\zeta})$ & $\boldsymbol{y}^{(\text{at-risk})}$ \\
M3 & $\Theta_{c,t} = \hat{\beta}^{\text{\tiny [M1]}}_c \eta_{c,t}\left(\boldsymbol{\hat{\phi}}^{\text{\tiny [M1]}} \right)$ & $\boldsymbol{y}^{(c)}$ \\
M4 & $\Theta_{c,t} = \hat{\beta}^{\text{\tiny [M1]}}_c \eta_{c,t}\left(\boldsymbol{\hat{\phi}}^{\text{\tiny [M1]}} \right) + \Omega_{c,t}\left(\boldsymbol{\hat{\zeta}}^{\text{\tiny [M2]}} \right)$ & $\boldsymbol{y}^{(c)}$ \\
\hline
\end{tabular}
\end{table}

For $c \in \mathcal{C}^{\text{inflation}}$, the projections under scenarios S1, S2 and S3 are also identical and based on posterior estimates of the model parameters from M2, combined with additional uncertainty related to the country baseline $\beta_c$ as obtained from M1:
\begin{equation}
^{(g)}{\Theta}^{\text{\tiny [S]}}_{c,t} =
^{(g)}{\beta}^{\text{\tiny [M1]}}_c
\cdot ^{(g)}{\eta}^{\text{\tiny [M2]}}_{c,t} +
^{(g)}{\delta}^{\text{\tiny [M2]}}_c \cdot 
^{(g)}{\Omega}^{\text{\tiny [M2]}}_{c,t},\text{ for scenario S}  \in \{\text{S1}, \text{S2}, \text{S3}\}, c \in \mathcal{C}^{\text{inflation}}.
\end{equation}

The construction of the scenario-based projections for the remaining countries, those at risk of inflation but without strong evidence of past or ongoing inflations, denoted by set $\mathcal{C}^{\text{future-inf}}$, differs between scenarios. In S2, the country-specific probability of having inflation, as obtained in M2, is used for constructing S2 projected trajectories $^{(g)}{\Theta}^{\text{\tiny [S2]}}_{c,t}$:
\begin{equation}
^{(g)}{\Theta}^{\text{\tiny [S2]}}_{c,t} =
^{(g)}{\beta}^{\text{\tiny [M1]}}_c
\cdot ^{(g)}{\eta}^{\text{\tiny [M2]}}_{c,t} +
^{(g)}{\delta}^{\text{\tiny [M2]}}_c \cdot 
^{(g)}{\Omega}^{\text{\tiny [M2]}}_{c,t}, \hspace{0.1cm}
c \in \mathcal{C}^{\text{future-inf}}. 
\end{equation}
 
For scenarios S1 and S3 for at-risk countries $c \in \mathcal{C}^{\text{future-inf}}$, additional model fits M3 and M4 (see Table~\ref{tab_modelfits}) are introduced to obtain projections based on estimates without inflation (S1) and with 100\% chance of inflation (S3). Projections for S1 are obtained as follows: 
\begin{eqnarray}
^{(g)}{\Theta}^{\text{\tiny [S1]}}_{c,t} &=& ^{(g)}{\beta}^{\text{\tiny [M1]}}_c \cdot ^{(g)}{\eta}^{\text{\tiny [M3]}}_{c,t}, \hspace{0.1cm}
c \in \mathcal{C}^{\text{future-inf}},
\end{eqnarray}
where the model equation for M3 corresponds to that from M1 - the  model without inflation terms - but M3 is fit to country-specific data $\boldsymbol{y}^{(c)}$, which includes \textit{all} data points from country $c$, as opposed to data up to 1970 only. By including all country-specific data in the model fit, resulting M3 estimates for recent years are informed by the country's data. Projections for S3 are given by:
\begin{eqnarray}
^{(g)}{\Theta}^{\text{\tiny [S3]}}_{c,t} &=& ^{(g)}{\beta}^{\text{\tiny [M1]}}_c \cdot ^{(g)}{\eta}^{\text{\tiny [M4]}}_{c,t} + ^{(g)}{\Omega}^{\text{\tiny [M4]}}_{c,t}, \hspace{0.1cm}
c \in \mathcal{C}^{\text{future-inf}},
\end{eqnarray}
where the model equation for M4 is the model with inflation term fit to country-specific data $\boldsymbol{y}^{(c)}$. The main difference between M2 and M4 is that the inflation $\delta_c =1$ for all trajectories in M4, hence forcing an inflation in all trajectories as per S3 assumption. 

We incorporate the uncertainty associated with the start year of the inflation, due to uncertainty in TFR projections, into the SRB projections for S2 and S3 in the projection result for the set of countries $\mathcal{C}^{\text{future-inf}}$. For trajectories with posterior samples of start year $\gamma_{0,c}$ beyond the most recent SRB observation, we add in additional uncertainty of $x_c$, the year in which TFR in country $c$ declines to 2.9, based on TFR trajectories. We use 1,000 TFR trajectories for associated uncertainty for each country-year based on projections from the \texttt{bayesTFR} \texttt{R}-package \cite{vsevvcikova2011bayestfr,r_bayestfr}.

\subsection{Model validation}
We assess model performance via validation exercises focused on: 1) predicting left-out recent SRB observations, and, 2) predicting SRB inflation transitions.

\subsubsection{Predicting left-out SRB observations}\label{sec-method-out-of-sample-vali}
We assess model performance by leaving out recent data. Specifically, we leave out 20\% of the data points that were collected after a certain survey year \cite{alkema2012monitoring,alkema2014national} for the out-of-sample validation. After leaving out data, we fitted the model to the training data set, and obtain median estimates and prediction intervals for the SRB and SRB inflation that would have been constructed based on available data set in the survey year selected. We also assess the model performance by leaving out data at random, i.e. leaving out 20\% of the data randomly, and repeat this exercise 30 times.

We calculate median errors and median absolute errors for the left-out SRB observations, where errors are defined as: $e_j = y_j - (\hat{y_j}|\boldsymbol{y}^\text{train})$, where $(\hat{y_j}|\boldsymbol{y}^\text{train})$ refers to the posterior median of the predictive distribution based on the training dataset $\boldsymbol{y}^\text{train}$ for the $j$-th left-out observation $y_j$. Coverage for 95\% prediction intervals for left-out observations is given by $1/J \cdot \sum_{j=1}^J \mathbb{I}_{\mathbf{A}}(y_j, l_j) \cdot \mathbb{I}_{\mathbf{B}}(y_j, u_j)$, where $J$ is the total number of left-out observations, $l_j$ and $u_j$ correspond to the 2.5th and 97.5th percentiles of the posterior predictive distribution (PPD) for the $j$-th left-out observation $y_j$, and sets $\mathbf{A} = \{(a,b) \in \mathbb{R}^2: a>b\}$ and $\mathbf{B} = \{(a,b) \in \mathbb{R}^2: a<b\}$. For the 80\% prediction interval coverage, $l_j$ and $u_j$ refer to the 10th and 90th percentiles of the PPD respectively. The validation measures are calculated for 1000 permutations of left-out observations, where each permutation consists one randomly selected left-out observation from each country with data left out. The reported validation results are based on the mean of the outcomes from the 1000 permuted sets of left-out observations.

For the median estimates based on full data set and training data set, errors are defined as $e(\Theta)_{c, t} =
(\widehat{\Theta}_{c,t}|\boldsymbol{y}^\text{full}) - (\widehat{\Theta}_{c,t}|\boldsymbol{y}^\text{train})$, where $(\widehat{\Theta}_{c,t}|\boldsymbol{y}^\text{full})$ is the posterior median for country $c$ in year $t$ based on the full dataset $\boldsymbol{y}^\text{full}$, and $(\widehat{\Theta}_{c,t}|\boldsymbol{y}^\text{train})$ is the posterior median for the same country-year based on the training dataset. Similarly, the error for the sex ratio transition process with probability is defined as $e(\delta\Omega)_{c, t} =
(\widehat{\delta_c\Omega_{c,t}}|\boldsymbol{y}^\text{full}) - (\widehat{\delta_c\Omega_{c,t}}|\boldsymbol{y}^\text{train})$. Coverage is computed in a similar manner as for the left-out observations, based on the lower and upper bounds of the equal-tail 95\% credible interval for $\Theta_{c,t}$ based on the training dataset.

\subsubsection{Predicting sex ratio transitions since 1970}\label{sec_method_predict}
We assess the predictive performance of the inflation model by predicting the SRB inflation for each at-risk country, referring to the set of countries $\mathcal{C}^{\text{inflation}}\cup \mathcal{C}^{\text{future-inf}}$. In this exercise, country data after 1970 is not used to directly inform the transition parameter estimates. Instead, we use median estimates of the baseline and natural fluctuations up to 1970 based on country-specific data prior to 1970 and predict the SRB from 1970 onwards as follows:
\begin{eqnarray*}
^{(g)}{\widetilde{\Theta}}_{c,t} &=&
^{(g)}{\beta}^{\text{\tiny [M1]}}_c \cdot ^{(g)}{\widetilde{\eta}}_{c,t}
+ ^{(g)}{\widetilde{\delta}}_c \cdot  ^{(g)}{\widetilde{\Omega}}_{c,t} (^{(g)}{\boldsymbol{\zeta}}^{\text{\tiny [M2]}}),
\end{eqnarray*}
where $^{(g)}{\widetilde{\delta}}_c$ and $^{(g)}{\widetilde{\Omega}}_{c,t}$ refer to samples from their respective posterior predictive distributions, where $^{(g)}{\widetilde{\delta}}_c$ is simulated based on $^{(g)}{\mu}^{\text{\tiny [M2]}}_{\pi}$ and $^{(g)}{\sigma}^{\text{\tiny [M2]}}_{\pi}$, and $^{(g)}{\widetilde{\Omega}}_{c,t}$ is simulated based on $^{(g)}{\boldsymbol{\zeta}}^{\text{\tiny [M2]}}$ where
\begin{equation*}
^{(g)}{\boldsymbol{\zeta}}^{\text{\tiny [M2]}} = \{ ^{(g)}{\mu}^{\text{\tiny [M2]}}_{\xi}, ^{(g)}{\sigma}^{\text{\tiny [M2]}}_{\xi}, ^{(g)}{\mu}^{\text{\tiny [M2]}}_{\lambda1}, ^{(g)}{\sigma}^{\text{\tiny [M2]}}_{\lambda1}, ^{(g)}{\mu}^{\text{\tiny [M2]}}_{\lambda2}, ^{(g)}{\sigma}^{\text{\tiny [M2]}}_{\lambda2}, ^{(g)}{\mu}^{\text{\tiny [M2]}}_{\lambda3}, ^{(g)}{\sigma}^{\text{\tiny [M2]}}_{\lambda3} \}.
\end{equation*}
The natural fluctuations $^{(g)}{\widetilde{\eta}}_{c,t}$ are given by M2 for $t \leq 1970$ and predicted for $t > 1970$:
\begin{eqnarray*}
^{(g)}{\widetilde{\eta}}_{c,t} &=& \cdot ^{(g)}{\eta}^{\text{\tiny [M2]}}_{c,t}, \text{ for } t \leq 1970, \\
\log(^{(g)}{\widetilde{\eta}}_{c,t}) | ^{(g)}{\widetilde{\eta}}_{c,t-1}&\sim&
\mathcal{N}\left(^{(g)}{\rho}^{\text{\tiny [M1]}} \log(^{(g)}{\widetilde{\eta}}_{c,t-1}), \left( ^{(g)}{\sigma}^{\text{\tiny [M1]}}_{\epsilon} \right)^2 \right), \text{for }t \in \{1971, \cdots, 2100\}.
\end{eqnarray*}

After generating the predicted values, we calculate the same set of results as described in Section~\ref{sec-method-out-of-sample-vali}.

\subsection{Estimates of missing female births}\label{sec_methodcmfb}
The realization of SRB inflation due to sex-selective abortion is quantified with the number of ``missing female births'', which refers to the additional number of female births that would have been born if the inflation were absent. It is calculated as the difference between the number of female births under normal circumstance (referred here as the ``inflation-free'' number of female births) and the actual number under the observed SRB.

The estimated and expected inflation-free number of female live births for a country-year, denoted by $\Psi_{c,t}$ and $\Psi_{c,t}^{(\text{inflation-free})}$, are computed as: 
\begin{eqnarray*}
\Psi_{c,t} &=& B_{c, t}/(1 + \Theta_{c,t}), \text{ and}\\
\Psi_{c,t}^{(\text{inflation-free})} &=& (B_{c, t} - \Psi_{c,t})/\Theta_{c,t}^{(\text{inflation-free})},
\end{eqnarray*}
where $B_{c,t}$ is the total number of births for a certain country-year, obtained from the UN WPP 2019 \cite{UNWPP2019}. The number of inflation-free female births $\Psi_{c,t}^{(\text{inflation-free})}$ is obtained from the estimated number of male births $(B_{c, t} - \Psi_{c,t})$, and the inflation-free SRB $\Theta_{c,t}^{(\text{inflation-free})} = \beta_c \eta_{c,t}$ for the respective country-year. The method of calculating the missing female births follows from what was first introduced by \cite{dreze1990hunger} and was recently reviewed and validated by \cite{guilmoto2020mis}.
 
The annual number of missing female births (AMFB) for country $c$ in year $t$ is defined as: $$\Psi_{c,t}^{(\text{missing})} = \Psi_{c,t}^{(\text{inflation-free})} - \Psi_{c,t}.$$ The cumulative number of missing female births (CMFB) for period $t_1$ to $t_2$ in country $c$ is defined as the sum of AMFB from the year $t_1$ up to the year $t_2$: $$\Lambda_{c, [t_1,t_2]}^{(\text{missing})} = \sum_{t = t_1}^{t_2}\Psi_{c,t}^{(\text{missing})}.$$

\section{Results}\label{sec_results}
We first summarize findings related to SRB baselines and natural deviations for country-years without risk of inflation. Secondly, the sex ratio transition model results are presented, followed by projections. Finally, validation results are presented. 

\subsection{SRB baselines and country-year variations in years without risk of inflation}
\begin{figure}
\centering
\includegraphics[width=\linewidth]{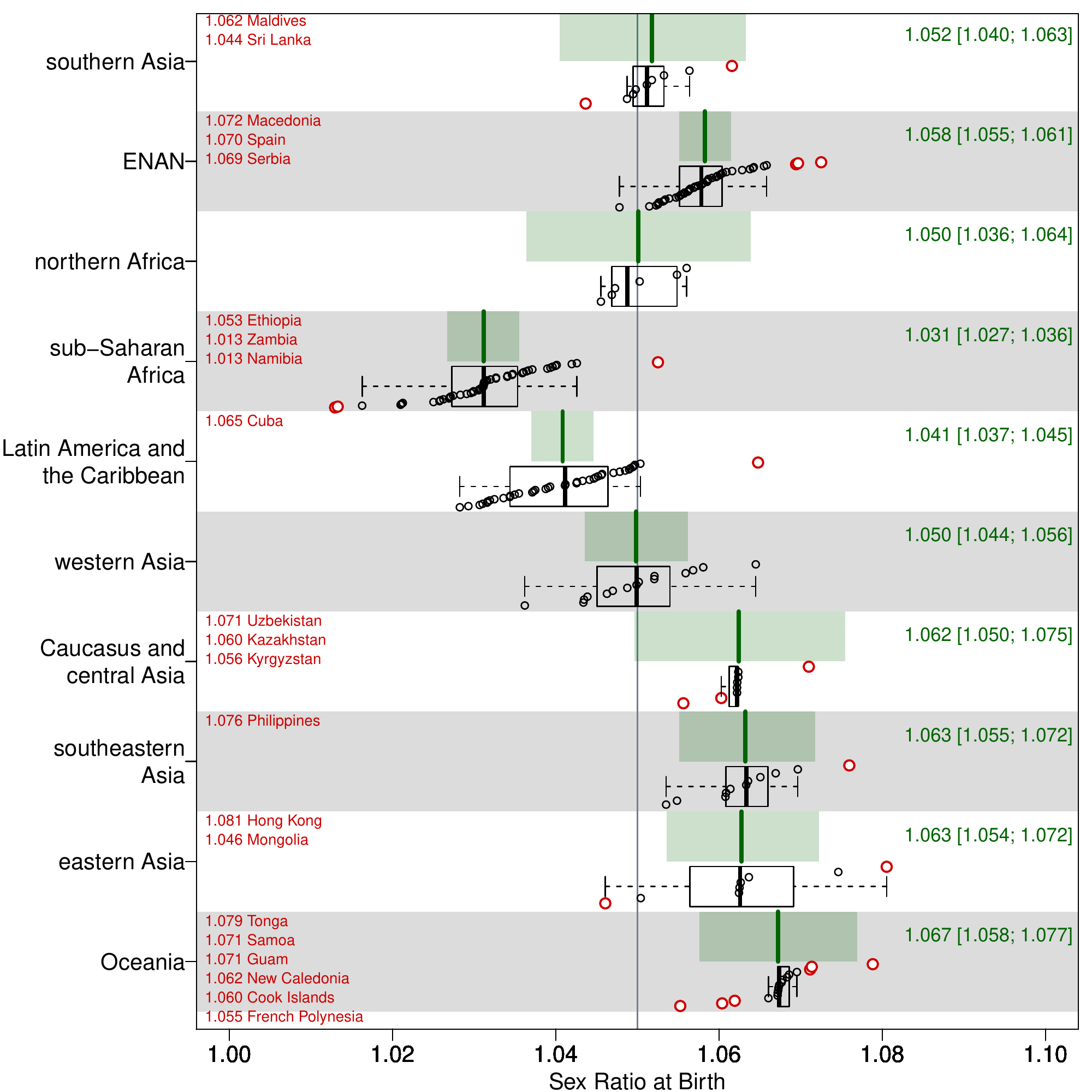}
\caption[SRB regional and national baseline]{\textbf{SRB regional and national baseline.} Regional baseline median estimates $\beta_r^{(\text{region})}$ (dark green line), with 95\% credible intervals (shaded area) and printed values (green). Box plots summarize the distributions of national baselines $\beta_c$ within each region. National median estimates are shown (dots) with those outside the range of [25th percentile - 1.5IQR; 75th percentile+1.5IQR] highlighted in red. The countries with outlying national baselines are listed in the legend with the median estimates reported (ordered by median estimate). ENAN: the combination of countries in Europe, North America, Australia, and New Zealand.}
\label{fig_baseline}
\end{figure}

The median estimates of regional SRB baselines $\beta_r^{(\text{region})}$ and national baselines $\beta_c$ are illustrated in Figure~\ref{fig_baseline}. The lowest regional baseline is estimated in sub-Saharan Africa at 1.031 (95\% credible interval [1.027; 1.036]) and the highest is in the Oceania at 1.067 [1.058; 1.077]. Among the ten regions, the probability that baseline is smaller than 1.05 (the widely assumed historical norm) is greater than 97.5\% in two regions (sub-Saharan Africa, Latin America and the Caribbean) and the probability that baseline is larger than 1.05 is bigger than 97.5\% in four regions (ENAN, southeastern Asia, eastern Asia, and Oceania). The national baselines are estimated to range from 1.013 [1.000; 1.026] in Zambia and 1.013 [0.997; 1.028] in Namibia to 1.081 [1.068; 1.093] in Hong Kong.

\begin{figure}
\centering
\includegraphics[width=0.8\linewidth]{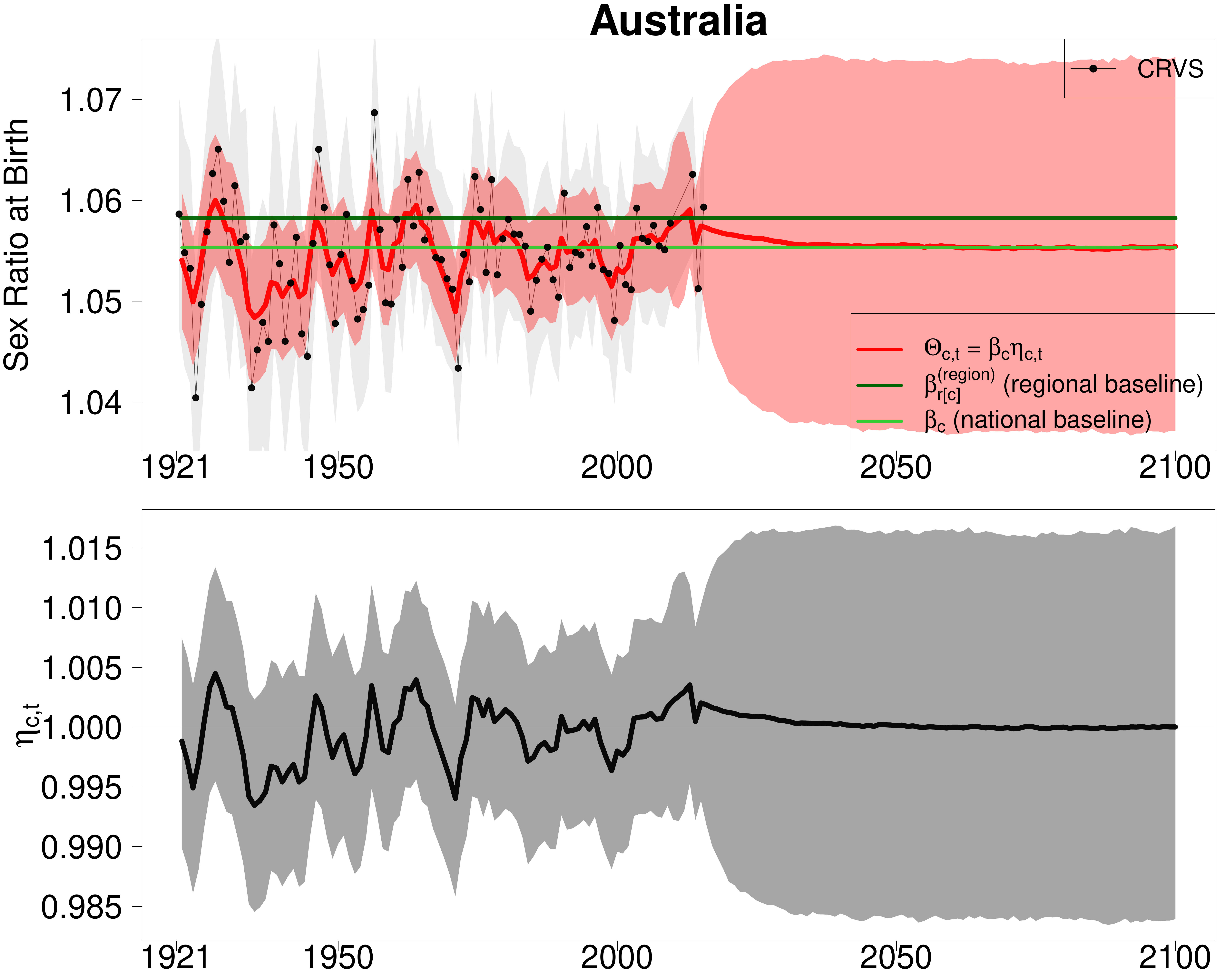}
\caption[SRB estimates and projections for Australia]{\textbf{SRB estimates and projections for Australia.} Top row: SRB median estimates $\Theta_{c,t}$ (red curve) and 95\% credible intervals (red shades), median estimates of the regional baselines $\beta_{r[c]}^{(\text{region})}$ (dark green horizontal line), median estimates of the national baselines $\beta_c$ (light green horizontal line). SRB observations are displayed with dots and observations are connected with lines when obtained from the same source. Shaded areas around observation series represent the sampling variability in the series (quantified by two times the stochastic/sampling standard errors). Bottom row: median estimates of natural deviation $\eta_{c,t}$ (solid line) and 95\% credible intervals (shades).}
\label{fig_s1_Australia}
\end{figure}

Australia is an example of a country without risk of SRB inflation. Its estimates and projections are displayed in Figure~\ref{fig_s1_Australia}. It typifies countries with high quality annual CRVS data, here available from 1921 to 2015. SRB estimates follow the CRVS data trend and the uncertainty assessment takes into account the stochastic uncertainty associated with the CRVS data. Its national baseline $\beta_c$ is estimated at 1.055 [1.049; 1.062]. The national baseline of Australia differs from its regional baseline $\beta_{r[c]}^{(\text{region})}$ at 1.058 [1.055; 1.061] for the region ENAN (the combination of countries in Europe, North America, Australia, and New Zealand) since the national baseline is informed by the CRVS data available in Australia. The estimated SRB for Australia $\Theta_{c,t}$ ranges from 1.048 [1.042; 1.055] in 1935 to 1.060 [1.054; 1.067] in 1928. As Australia is not identified to have SRB inflation risk, all the deviations of $\Theta_{c,t}$ from the national baseline of SRB $\beta_c$ are considered as natural fluctuation and are captured with $\eta_{c,t}$ (Figure~\ref{fig_s1_Australia}, bottom row). Given that Australia is considered as not having risk of SRB inflation, its projected SRB is based on the model of country-years without SRB inflation. The SRB projection for Australia is approximately constant and given by its national baseline $\beta_c$, with the projection for 2100 given by 1.055 [1.037; 1.074].

\subsection{Sex ratio transition model}
Table~\ref{table_sup_allADJ} summarize model results for the sex ratio transition for the 29 countries with risk of SRB inflation. In total, 12 countries are identified as having strong statistical evidence of SRB inflation: Albania; Armenia; Azerbaijan; China; Georgia; Hong Kong, SAR of China; India; Republic of Korea; Montenegro; Taiwan, Province of China; Tunisia; Vietnam. 

\begin{figure}
\includegraphics[width=\linewidth]{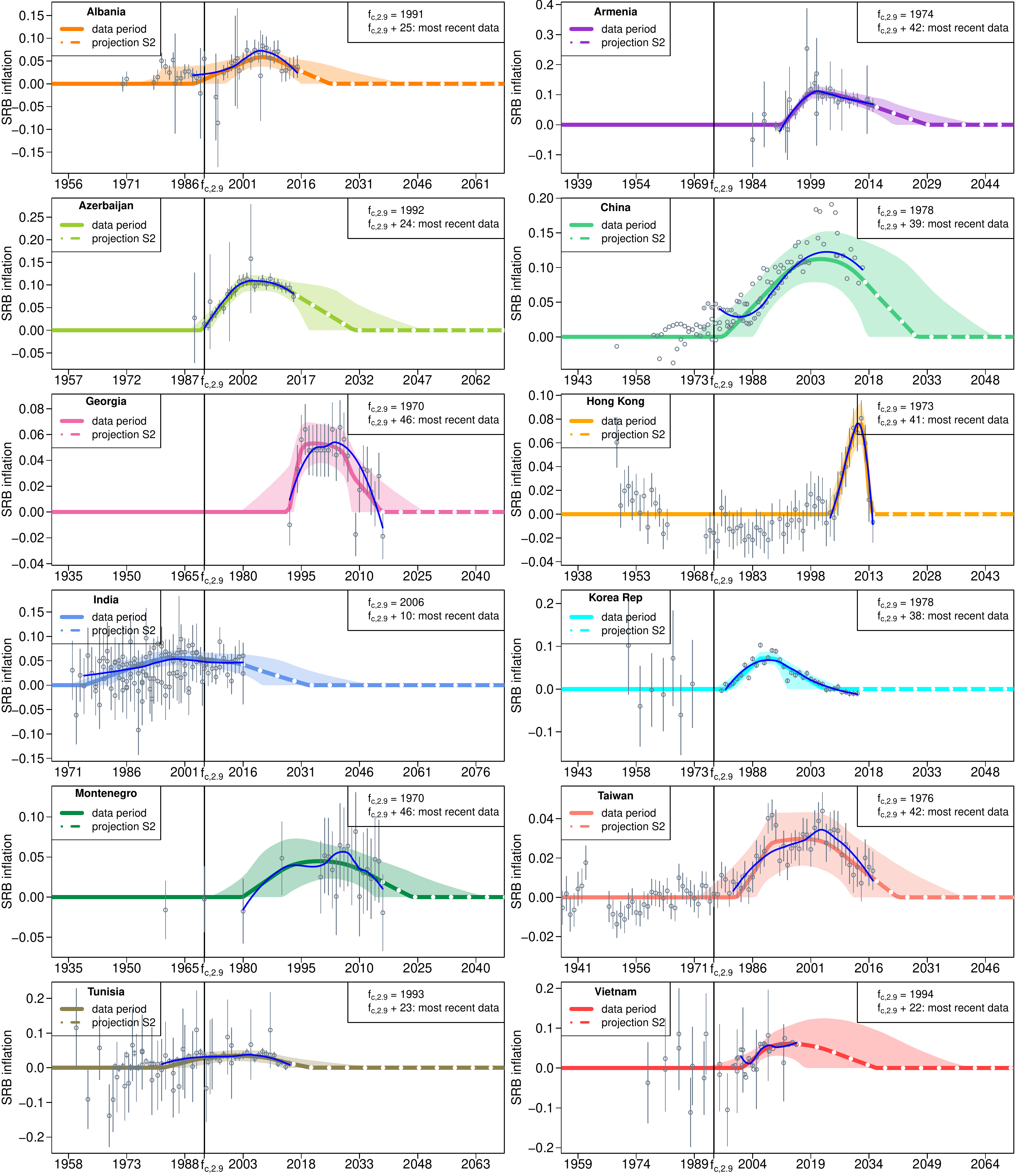}
\caption[SRB inflation estimates and projections for countries with strong statistical evidence of SRB inflation, with inflation residuals]{\textbf{SRB inflation estimates and projections for countries with strong statistical evidence of SRB inflation, with inflation residuals.} SRB inflation median estimates during periods with data (solid lines), median projections (dashed lines), and 95\% credible intervals (shades). Data shown are the difference between the observed SRB and the median estimates of country-specific baseline (i.e. $y_i - \hat{\beta_c}^{\text{\tiny [M1]}}$). Vertical line segments around data represent the sampling variability in the data (quantified by two times the stochastic/sampling standard errors). Loess curves are in blue. Loess curves are based on data after the inflation start year median estimates and data points are weighted by the inverse of sampling variance. The year in which the TFR declines to 2.9 in each country is shown as well (vertical line labeled as $f_{c,2.9}$ on the x-axis).}
\label{fig_srb_inflation}
\end{figure}

Figure~\ref{fig_srb_inflation} presents the estimated SRB inflation $\Omega_{c,t}$ (rather than SRB itself) together with inflation residuals for the 12 countries with strong evidence of SRB inflation. We obtain the difference between the observed SRBs and the median estimates of country-specific baselines, referred to here as inflation residual $r_i = y_i - \hat{\beta_c}^{\text{\tiny [M1]}}$.
The residuals indicate the presence of an inflation and the parametric form of $\Omega_{c,t}$  is able to capture its shape. 
For the 12 countries, the median estimate of start year of the sex ratio transition $\gamma_{0,c}$ is estimated to be the year in which TFR declines to 5.2 in India to 1.0 in Hong Kong (SAR of China). The country-specific maximum value of SRB imbalance $\xi_c$ is estimated to be higher than 0.100 in China at 0.112 [0.078; 0.151], in Armenia at 0.108 [0.088; 0.129] and in Azerbaijan at 0.104 [0.087; 0.121]. The maximum imbalance is lower than 0.050 in Montenegro at 0.045 [0.022; 0.068], in Tunisia at 0.035 [0.019; 0.050] and in Taiwan (Province of China) at 0.030 [0.016; 0.043]. The estimated duration of the sex ratio transition, referring to the sum of the 3 phases $\lambda_{1,c}+\lambda_{2,c}+\lambda_{3,c}$, varies greatly across the 12 countries and ranges from relatively fast transitions that last 11, 25 and 26 years for Hong Kong (SAR of China), Georgia, and the Republic of Korea respectively, to 45, 51 and 58 years in Montenegro, China and India. For the 12 countries, the average projection period from the most recent observation year to end of transition is 8.8 years, corresponding to an average of 20.1\% of country-specific duration of the sex ratio inflation. All these countries have data available up to the 2nd phase of the sex ratio transition, the stagnation. At the country level, transitions are estimated to have finished prior to the most recent observation year in Hong Kong (SAR of China), and Republic of Korea. Four countries (Georgia, Montenegro, Taiwan (Province of China), and Tunisia) are estimated to be in transition phase 3 and the remaining six countries (Albania, Armenia, Azerbaijan, China, India, and Vietnam) are estimated to be in phase 2.

\begin{figure}
\centering
\includegraphics[width=0.8\linewidth]{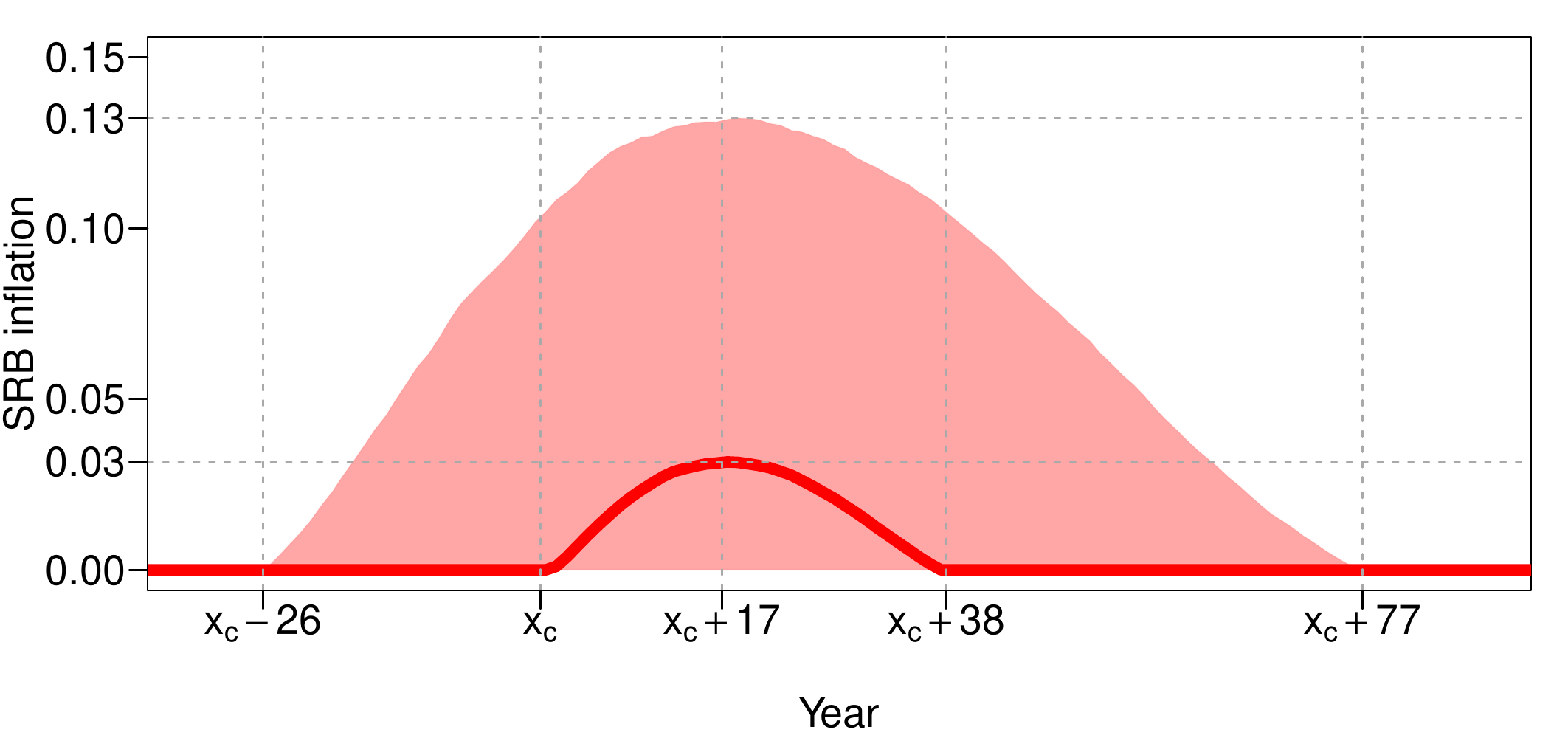}
\caption[SRB inflation projection for a country with future SRB inflation/prior to observing data]{\textbf{SRB inflation projection for a country with future SRB inflation/prior to observing data.} Median projection (solid line) with 95\% credible intervals (shades). $x_c$ refers to the year in which the total fertility rate (TFR) in a country declines to 2.9 or 1970, whichever occurs later.}
\label{fig_srb_inflation_sim}
\end{figure}

Figure~\ref{fig_srb_inflation_sim} presents the projected SRB inflation $\Omega_{c,t}$ for a country prior to observing SRB data. Given the hierarchical structure of the SRB inflation model, the sex ratio transition in Figure~\ref{fig_srb_inflation_sim} represents the average experience of SRB imbalance. The median projected SRB inflation process ($\lambda_{1,c}+\lambda_{2,c}+\lambda_{3,c}$) has a span of 37 [15; 64] years. The maximum SRB inflation $\xi_c$ for a new country has a median at 0.032 with 95\% credible interval [0.000; 0.132]. The inflation maximum is reached around 11 [1; 28] years after the country's TFR declines to 2.9 in the year $x_c$.

% latex table generated in R 3.6.1 by xtable 1.8-4 package
% Mon May 25 10:58:22 2020
\begingroup\footnotesize
\begin{longtable}{|p{1.7cm}|c|c|c|c|c|c|c|c|c|}
\caption[Sex ratio transition model results for the 29 countries at risk of SRB inflation]{\textbf{Sex ratio transition model results for the 29 countries at risk of SRB inflation.} Numbers above the brackets are posterior median estimates. Numbers inside the brackets are 95\% credible intervals. $\psi_c$: the relative inclusion of SRB inflation. $\gamma_{0,c}$: the start year of SRB inflation. $\lambda_{1,c}$, $\lambda_{2,c}$, $\lambda_{3,c}$: period lengths for increase, stagnation and decrease for the sex ratio transition process. $\gamma_{3,c}=\gamma_{0,c}+\lambda_{1,c}+\lambda_{2,c}+\lambda_{3,c}$: the end year of SRB inflation. $\xi_c$: maximum SRB inflation. 2100+: indicates the year is beyond 2100. Countries with strong statistical evidence of SRB inflation are in boldface.}
\label{table_sup_allADJ}\\
\hline Country  & $\psi_c$ & TFR in & $\gamma_{0,c}$ & $\gamma_{3,c}$ & $\xi_c$ & $\delta_c\xi_c$& $\lambda_{1,c}$ & $\lambda_{2,c}$ & $\lambda_{3,c}$  \\
 & (in \%) & $\gamma_{0,c}$ & (start year) & (end year) &  &  & & &\\\hline
\endfirsthead
\multicolumn{10}{c}
{{\bfseries \tablename\ \thetable{} -- continued from previous page}} \\
\hline Country  & $\psi_c$ & TFR in & $\gamma_{0,c}$ & $\gamma_{3,c}$ & $\xi_c$ & $\delta_c\xi_c$ & $\lambda_{1,c}$ & $\lambda_{2,c}$ & $\lambda_{3,c}$ \\ 
& (in \%) & $\gamma_{0,c}$ & (start year) & (end year) &  &  &&&\\\hline\endhead
\multicolumn{10}{r}{{Continued on next page}} \\
\endfoot
\endlastfoot
\textbf{Albania} & 100 & 3.1 & 1988 & 2024 & 0.059  & 0.059 & 15 & 3 & 15 \\ 
 &  &  & [1973; 1997] & [2016; 2043] & [0.038; 0.079] & [0.038; 0.079] & [3; 31] & [0; 26] & [6; 33]\\ 
   \hline
\textbf{Armenia} & 100 & 2.5 & 1992 & 2029 & 0.109 & 0.109 & 7 & 5 & 26\\ 
   &  &  & [1990; 1993] & [2020; 2042] & [0.088; 0.13] & [0.088; 0.130] & [5; 9] & [0; 19] & [7; 39]\\ 
   \hline
\textbf{Azerbaijan} & 100 & 3.0 & 1991 & 2031 & 0.104  & 0.104 & 10 & 9 & 19\\ 
   &  &  & [1988; 1993] & [2019; 2049] & [0.087; 0.122] & [0.087; 0.122] & [7; 15] & [2; 23] & [3; 39] \\ 
   \hline
\textbf{China} & 100 & 2.6 & 1980 & 2030 & 0.114 & 0.114& 20 & 11 & 18 \\ 
   &  &  & [1972; 1988] & [2017; 2051] & [0.08; 0.156] & [0.080; 0.156] & [8; 32] & [1; 28] & [3; 39] \\ 
   \hline
\textbf{Georgia} & 100 & 2.1 & 1992 & 2016 & 0.054 & 0.054 & 3 & 9 & 12 \\
   &  &  & [1979; 1994] & [2008; 2027] & [0.039; 0.07] & [0.039; 0.070] & [0; 20] & [1; 13] & [0; 27] \\ 
   \hline
\textbf{Hong Kong,} & 100 & 1.0 & 2004 & 2013 & 0.076 & 0.076 & 6 & 2 & 2 \\
\textbf{SAR of China}  &  &  & [2002; 2005] & [2012; 2014] & [0.059; 0.096] & [0.059; 0.096] & [4; 9] & [0; 4] & [0; 3] \\ 
   \hline
\textbf{India} & 100 & 5.2 & 1975 & 2033 & 0.056 & 0.056 & 19 & 16 & 24 \\
   &  &  & [1970; 1981] & [2021; 2050] & [0.042; 0.072] & [0.042; 0.072] & [7; 29] & [2; 34] & [3; 44] \\ 
   \hline
\textbf{Republic of} & 100 & 2.4 & 1982 & 2006 & 0.072 & 0.072 & 8 & 4 & 13 \\
\textbf{Korea}   &  &  & [1978; 1984] & [1997; 2011] & [0.058; 0.087] & [0.058; 0.087] & [6; 12] & [2; 5] & [2; 18] \\ 
   \hline
\textbf{Tunisia} & 100 & 4.9 & 1982 & 2021 & 0.036 & 0.036 & 13 & 10 & 15 \\
   &  &  & [1976; 1989] & [2012; 2039] & [0.021; 0.052] & [0.021; 0.052] & [3; 25] & [1; 21] & [2; 36] \\ 
   \hline
\textbf{Vietnam} & 100 & 2.0 & 2001 & 2036 & 0.066 & 0.066 & 10 & 8 & 16 \\
   &  &  & [1991; 2005] & [2017; 2061] & [0.035; 0.131] & [0.035; 0.131] & [1; 26] & [0; 25] & [2; 37] \\ 
   \hline
\textbf{Montenegro} & 100 & 2.3 & 1980 & 2024 & 0.048 & 0.048 & 14 & 10 & 19 \\
   &  &  & [1971; 1990] & [2014; 2043] & [0.026; 0.076] & [0.026; 0.076] & [2; 29] & [1; 26] & [3; 39] \\ 
   \hline
\textbf{Taiwan,} & 99.7 & 2.4 & 1982 & 2023 & 0.031 & 0.031 & 10 & 12 & 20 \\
\textbf{Province of China}   &  &  & [1972; 1987] & [2012; 2041] & [0.018; 0.045] & [0.017; 0.044] & [3; 27] & [1; 22] & [5; 39] \\ 
   \hline
Mauritania & 63.8 & 2.9 & 2065 & 2100+ & 0.064 & 0.034 & 11 & 8 & 16 \\
   &  &  & [2038; 2093] & [2068; 2100+] & [0.005; 0.163] & [0.000; 0.149] & [1; 28] & [0; 25] & [1; 37] \\ 
   \hline
Mali & 63.4 & 2.9 & 2061 & 2099 & 0.064 & 0.033 & 11 & 8 & 16 \\
   &  &  & [2036; 2089] & [2066; 2100+] & [0.005; 0.16] & [0.000; 0.146] & [1; 28] & [0; 25] & [1; 37] \\ 
   \hline
Afghanistan & 63.3 & 2.9 & 2033 & 2071 & 0.063 & 0.033 & 11 & 8 & 16 \\
   &  &  & [2013; 2063] & [2041; 2100+] & [0.005; 0.16] & [0.000; 0.146] & [1; 28] & [0; 25] & [1; 37] \\ 
   \hline
Nigeria & 63.2 & 2.9 & 2065 & 2100+ & 0.064 & 0.033 & 11 & 8 & 16 \\
   &  &  & [2038; 2093] & [2069; 2100+] & [0.005; 0.163] & [0.000; 0.148] & [1; 28] & [0; 25] & [1; 37] \\ 
   \hline
Gambia & 63.1 & 2.9 & 2053 & 2091 & 0.064 & 0.033 & 11 & 8 & 16 \\
   &  &  & [2027; 2082] & [2057; 2100+] & [0.005; 0.161] & [0.000; 0.149] & [1; 28] & [0; 25] & [1; 37] \\ 
   \hline
Pakistan & 63.0 & 2.9 & 2030 & 2068 & 0.062 & 0.030 & 11 & 8 & 16 \\
   &  &  & [1995; 2058] & [2027; 2100+] & [0.005; 0.16] & [0.000; 0.146] & [1; 28] & [0; 25] & [1; 37] \\ 
   \hline
Senegal & 63.0 & 2.9 & 2061 & 2099 & 0.064 & 0.033 & 11 & 8 & 16 \\
   &  &  & [2034; 2089] & [2064; 2100+] & [0.005; 0.161] & [0.000; 0.148] & [1; 28] & [0; 25] & [1; 37] \\ 
   \hline
Tanzania & 62.7 & 2.9 & 2068 & 2100+ & 0.064 & 0.033 & 11 & 8 & 16 \\
   &  &  & [2041; 2096] & [2071; 2100+] & [0.005; 0.162] & [0.000; 0.146] & [1; 28] & [0; 25] & [1; 37] \\ 
   \hline
Uganda & 62.6 & 2.9 & 2042 & 2080 & 0.064 & 0.032 & 11 & 8 & 16 \\
   &  &  & [2020; 2070] & [2049; 2100+] & [0.005; 0.162] & [0.000; 0.147] & [1; 28] & [0; 25] & [2; 37] \\ 
   \hline
Nepal & 62.6 & 2.6 & 2009 & 2047 & 0.058 & 0.026 & 12 & 8 & 16 \\
   &  &  & [1989; 2036] & [2018; 2083] & [0.005; 0.155] & [0.000; 0.135] & [1; 28] & [0; 25] & [1; 37] \\ 
   \hline
Tajikistan & 62.2 & 2.9 & 2038 & 2076 & 0.063 & 0.032 & 11 & 8 & 16 \\
   &  &  & [2016; 2067] & [2044; 2100+] & [0.005; 0.162] & [0.000; 0.147] & [1; 28] & [0; 25] & [1; 37] \\ 
   \hline
Egypt & 61.9 & 2.9 & 2030 & 2068 & 0.063 & 0.030 & 11 & 8 & 16 \\
   &  &  & [2006; 2058] & [2034; 2100+] & [0.005; 0.161] & [0.000; 0.147] & [1; 28] & [0; 25] & [1; 37] \\ 
   \hline
Jordan & 56.3 & 2.7 & 2019 & 2057 & 0.061 & 0.017 & 11 & 8 & 16 \\
   &  &  & [2000; 2048] & [2028; 2095] & [0.004; 0.158] & [0.000; 0.140] & [1; 28] & [0; 25] & [1; 37] \\ 
   \hline
Singapore & 44.9 & 2.3 & 1975 & 2014 & 0.029 & 0.000 & 10 & 8 & 17 \\
   &  &  & [1970; 2012] & [1990; 2055] & [0.003; 0.145] & [0.000; 0.031] & [1; 27] & [0; 25] & [2; 37] \\ 
   \hline
Morocco & 40.8 & 2.6 & 2003 & 2041 & 0.061 & 0.000 & 11 & 8 & 16 \\
   &  &  & [1982; 2033] & [2010; 2080] & [0.003; 0.156] & [0.000; 0.127] & [1; 28] & [0; 25] & [1; 37] \\ 
   \hline
Bangladesh & 37.4 & 2.4 & 2009 & 2046 & 0.061 & 0.000 & 11 & 8 & 16 \\
   &  &  & [1987; 2038] & [2015; 2085] & [0.003; 0.162] & [0.000; 0.130] & [1; 28] & [0; 25] & [1; 37] \\ 
   \hline
Turkey & 34.9 & 2.9 & 1993 & 2031 & 0.046 & 0.000 & 11 & 7 & 16 \\
   &  &  & [1974; 2028] & [2001; 2074] & [0.003; 0.151] & [0.000; 0.078] & [1; 28] & [0; 25] & [2; 37] \\ 
  \hline
\end{longtable}
\endgroup

\clearpage

\subsection{SRB estimates and projections for countries at risk of SRB inflation}
SRB and resulting missing female births projections by scenario for all countries at risk of SRB inflation are given in the Figure~\ref{sup_fig_scenario_proj} (i.e. Supplementary figure A). Scenario 1 SRB estimates and projections during 1950--2100 for all countries are given in the Figure~\ref{fig_allcountry_s1} (i.e. Supplementary figure B). Here we illustrate projections for selected countries. 

\begin{figure}
\centering
\includegraphics[width=0.73\linewidth]{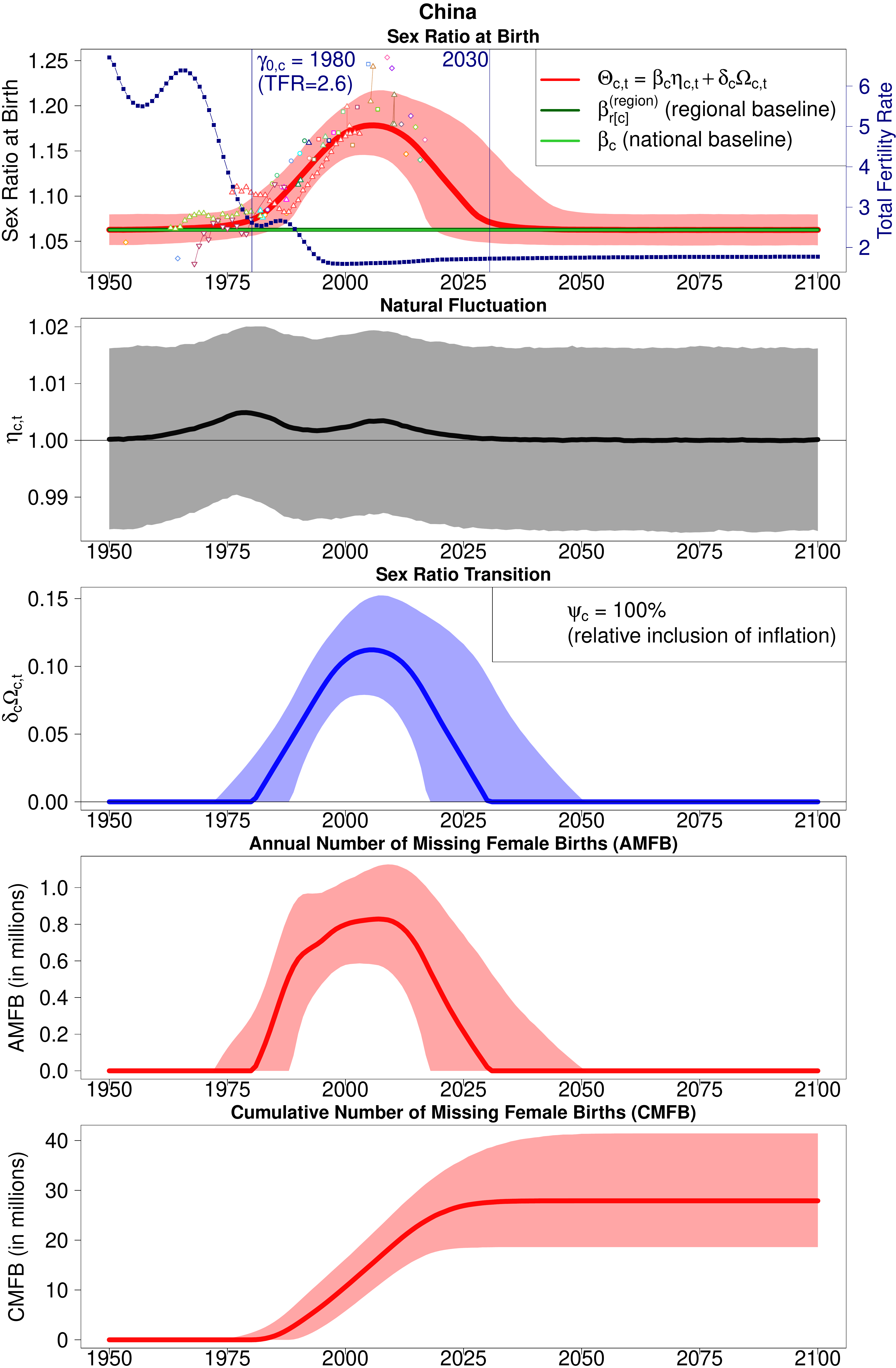}
\caption[SRB and missing female births estimates and projections for China]{\textbf{SRB and missing female births estimates and projections for China.} Row 1: SRB median estimates $\Theta_{c,t}$ (red curve) and 95\% credible intervals (red shades), median estimates of the regional baselines $\beta_{r[c]}^{(\text{region})}$ (dark green horizontal line), median estimates of the national baselines $\beta_c$ (light green horizontal line). SRB observations are displayed with dots and observations are connected with lines when obtained from the same source. Shaded areas around observation series represent the sampling variability in the series (quantified by two times the stochastic/sampling standard errors). The median estimates and projections of total fertility rate (TFR) from the UN WPP 2019 are added (blue squared dots). The median estimates of inflation start year $\gamma_{0,c}$ and end year $\gamma_{3,c}$ are the vertical lines. The TFR value in the year $\gamma_{0,c}$ is shown. Row 2: median estimates of natural deviation $\eta_{c,t}$ (solid line) and 95\% credible intervals (shades). Row 3: median estimates of SRB inflation $\delta_c\Omega_{c,t}$  (curves) and 95\% credible intervals (shades). Row 4: annual number of missing female births (AMFB) estimates and projections. Row 5: cumulative number of missing female births (CMFB) estimates and projections.}
\label{fig_s2_china}
\end{figure}

\paragraph*{Countries with strong statistical evidence of SRB inflation: China}
China is identified with strong statistical evidence of SRB inflation as listed in Table~\ref{table_sup_allADJ}. Its SRB inflation is still ongoing at the start of the projection period as shown in Figure~\ref{fig_s2_china}. The start year of the inflation is estimates at 1980 [1972; 1988] when the TFR declined to 2.6. We estimate that the SRB in China peaked in 2005 at 1.179 [1.141; 1.220], with associated SRB inflation of 0.112 [0.079; 0.150]. In 2017, the SRB in China is estimated at 1.144 [1.079; 1.206]. We project that the SRB will converge back to the range of natural fluctuations around its national baseline value of 1.063 [1.044; 1.082] in 2030 [2017; 2051].

The annual number of missing female births (AMFB) in China peaked in 2007 at 0.8 [0.6; 1.1] million female births per year. With the inflation projected to decrease to zero by 2030, the resulting AMFB is also projected to decrease to zero at that time. The cumulative number of missing female births (CMFB) for China since 1970 is projected to be 27.9 [18.6; 41.4] million by 2100.

\begin{figure}
\centering
\includegraphics[width=0.73\linewidth]{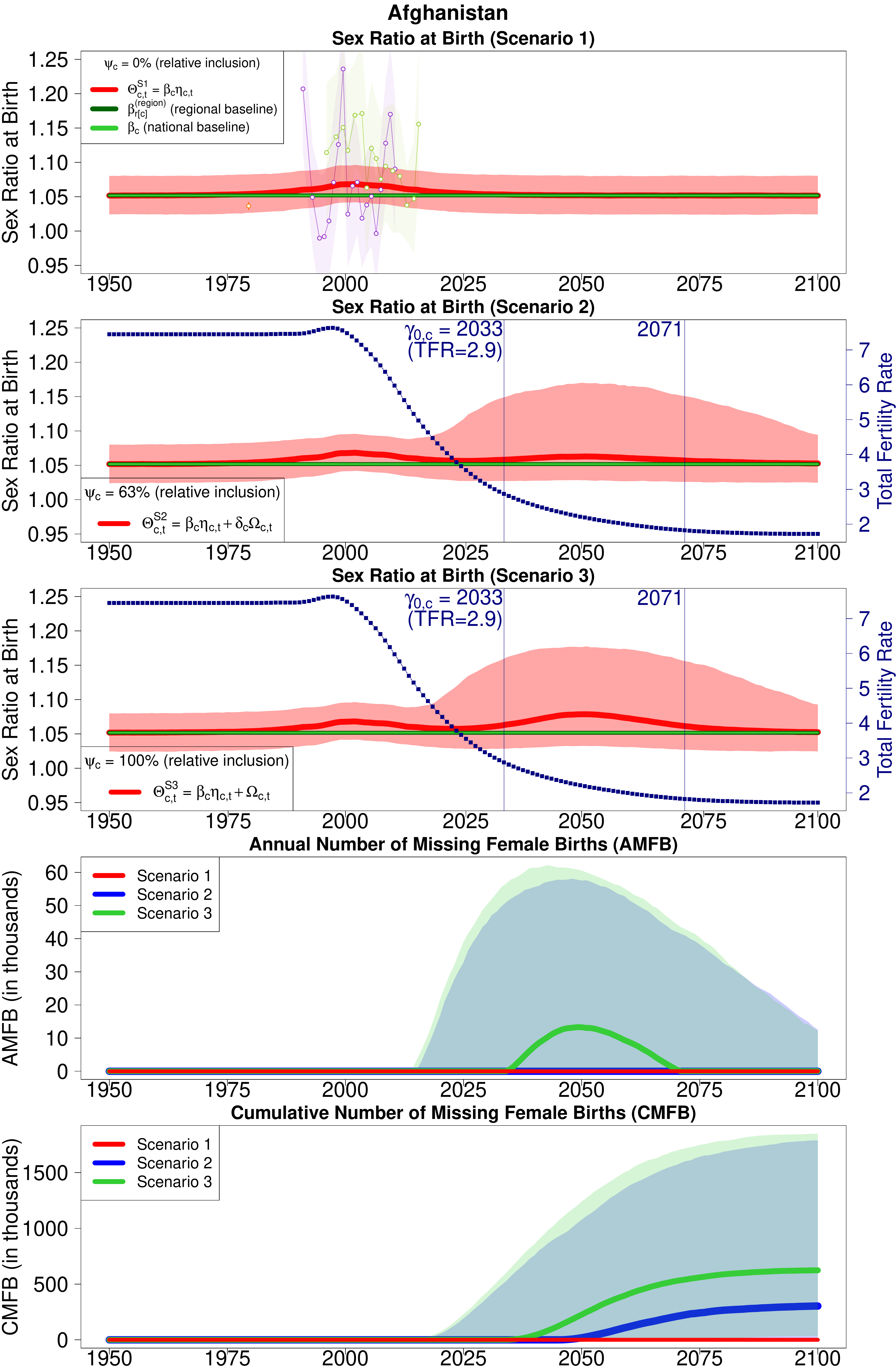}
\caption[SRB and missing female births estimates and scenario-based projections for Afghanistan]{\textbf{SRB and missing female births estimates and scenario-based projections for Afghanistan.} Row 1: SRB median estimates/projections and 95\% credible intervals for Scenario 1 $\Theta_{c,t}^{\text{\tiny [S1]}}$ (red curve and shades), median estimates of the regional baselines $\beta_{r[c]}^{(\text{region})}$ (dark green horizontal line), median estimates of the national baselines $\beta_c$ (light green horizontal line). SRB observations are displayed with dots and observations are connected with lines when obtained from the same source. Shaded areas around observation series represent the sampling variability in the series (quantified by two times the stochastic/sampling standard errors). Row 2: SRB median estimates/projections and 95\% credible intervals for Scenario 2 $\Theta_{c,t}^{\text{\tiny [S2]}}$ (curve and shades). Row 3: SRB median estimates/projections and 95\% credible intervals for Scenario 3 $\Theta_{c,t}^{\text{\tiny [S3]}}$ (curve and shades). The median estimates and projections of total fertility rate (TFR) from the UN WPP 2019 are added (blue squared dots). The median estimates of inflation start year $\gamma_{0,c}$ and end year $\gamma_{3,c}$ are the vertical lines. The TFR value in the year $\gamma_{0,c}$ is shown. Row 4: annual number of missing female births (AMFB) estimates and projections by scenario. Row 5: cumulative number of missing female births (CMFB) estimates and projections by scenario.}
\label{fig_s123_afg}
\end{figure}

\begin{figure}
\centering
\includegraphics[width=0.73\linewidth]{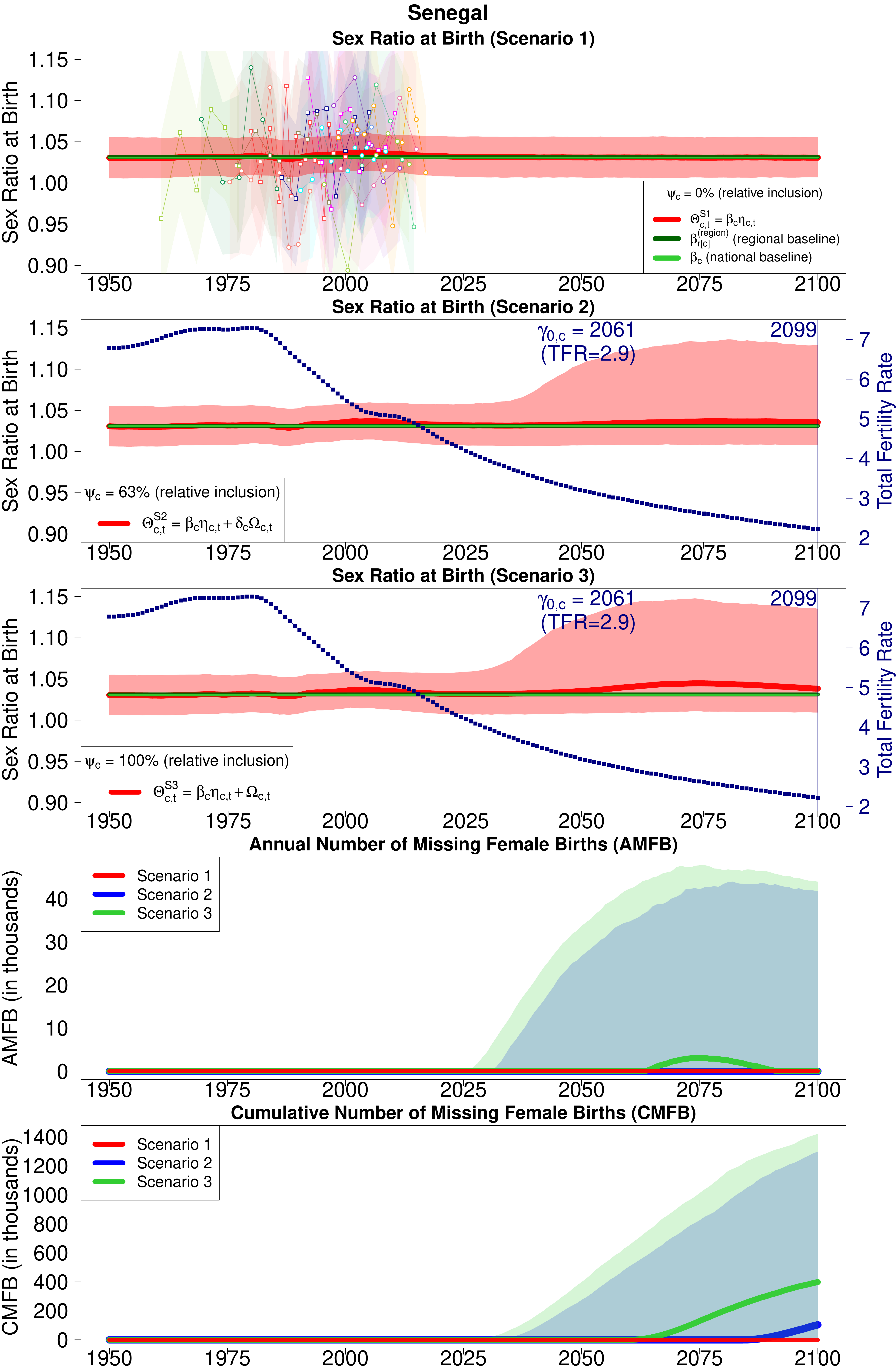}
\caption[SRB and missing female births estimates and scenario-based projections for Senegal]{\textbf{SRB and missing female births estimates and scenario-based projections for Senegal.} Row 1: SRB median estimates/projections and 95\% credible intervals for Scenario 1 $\Theta_{c,t}^{\text{\tiny [S1]}}$ (red curve and shades), median estimates of the regional baselines $\beta_{r[c]}^{(\text{region})}$ (dark green horizontal line), median estimates of the national baselines $\beta_c$ (light green horizontal line). SRB observations are displayed with dots and observations are connected with lines when obtained from the same source. Shaded areas around observation series represent the sampling variability in the series (quantified by two times the stochastic/sampling standard errors). Row 2: SRB median estimates/projections and 95\% credible intervals for Scenario 2 $\Theta_{c,t}^{\text{\tiny [S2]}}$ (curve and shades). Row 3: SRB median estimates/projections and 95\% credible intervals for Scenario 3 $\Theta_{c,t}^{\text{\tiny [S3]}}$ (curve and shades). The median estimates and projections of total fertility rate (TFR) from the UN WPP 2019 are added (blue squared dots). The median estimates of inflation start year $\gamma_{0,c}$ and end year $\gamma_{3,c}$ are the vertical lines. The TFR value in the year $\gamma_{0,c}$ is shown. Row 4: annual number of missing female births (AMFB) estimates and projections by scenario. Row 5: cumulative number of missing female births (CMFB) estimates and projections by scenario.}
\label{fig_s123_sen}
\end{figure}

\paragraph*{Countries at risk but without strong evidence of SRB inflation: Afghanistan and Senegal}
17 countries are classified with risk of SRB inflation without strong statistical evidence of SRB inflation as listed in Table~\ref{table_sup_allADJ}. The SRB in these countries are projected under three scenarios based on different assumptions on the occurrence of a sex ratio transition. We use Afghanistan and Senegal as examples to illustrate the scenario-based projections of SRB for such country (Figure~\ref{fig_s123_afg} and Figure~\ref{fig_s123_sen}). 

During the observation period for Afghanistan (Figure~\ref{fig_s123_afg}), data series for Afghanistan do not imply SRB inflation and hence all the fluctuations and deviations of $\Theta_{c,t}$ away from national baseline $\beta_c$ are captured with $\eta_{c,t}$. For Afghanistan, $\eta_{c,t}$ is close to one throughout the estimation period.

The three SRB projection scenarios result in substantively different future SRB and associated missing births in Afghanistan. The projection under S1 without future inflation, $\Theta_{c,t}^{\text{\tiny [S1]}}$, remains at its national baseline $\beta_c$. The S2 projection $\Theta_{c,t}^{\text{\tiny [S2]}}$ include a sex ratio transition in $\psi_c=63\%$ of all future trajectories while S3 projection $\Theta_{c,t}^{\text{\tiny [S3]}}$ includes transitions for all trajectories. For future trajectories with sex imbalances in Afghanistan under scenarios S2 and S3, the transition is projected to start in the 2030s ($\gamma_{0,c}$ is projected in the year 2033 [2013; 2063]) and ends in 2071 [2041; beyond 2100]. Given that we incorporate the uncertainty in the TFR for S2 and S3 projections $\Theta_{c,t}^{\text{\tiny [S2]}}$ and $\Theta_{c,t}^{\text{\tiny [S3]}}$, the SRB inflation process is more flattened than the one shown in Figure~\ref{fig_srb_inflation_sim}. Under scenario S2, the average annual number of missing female births (AMFB) during 2018--2100 is projected to be 4 [0; 22] thousand. The corresponding cumulative number of missing female births (CMFB) during 2018--2100 is projected to be 303 [0; 1787] thousand. Under scenario S3, the average AMFB is projected at 8 [0; 22] thousand during 2018--2100 and end up with 624 [36; 1848] thousand missing female births cumulatively.

For Senegal (Figure~\ref{fig_s123_sen}), the model detects no SRB inflation during the data period and projects the SRB imbalance to start in 2061 [2034; 2089] when the TFR declines to 2.9 and to end in 2099 [2064; beyond 2100]. The timing of the inflation is projected to be later than the sex ratio transition in Afghanistan due to differences in TFR projections, with Senegal's TFR projections suggesting a slower decline than expected in Afghanistan. For Senegal under scenario S2, the average AMFB from 2018 to 2100 is projected to be 1 [0; 16] thousand and corresponding CMFB is projected to be 104 [0; 1,298] thousand. The average AMFB and resulting CMFB for scenario S3 are 5 [0; 17] thousand and 399 [0; 1,422] thousand.

\subsection{Validation results}\label{sec_vali_res}
Table~\ref{tb-val-res-ppd} summarizes the results related to the left-out SRB observations for: (i) M1 (model without inflation as described in Section~\ref{sec_method_step2}) by leaving out observations obtained from the year 2005 onward, consisting 20.3\% of all observations with no risk of SRB inflation $\boldsymbol{y}^{(\text{risk-free})}$; (ii) M1 by  randomly leaving out 20\% observations (repeated 30 times); (iii) M2 (model with inflation term as described in Section~\ref{sec_method_step3}) by leaving out observations obtained from the year 2010 onward, consisting 19.8\% of the total observations at risk of inflation $\boldsymbol{y}^{(\text{at-risk})}$; and (iv) prediction of sex ratio transitions based on the prediction setup as described in Section~\ref{sec_method_predict}. Median errors and median absolute errors are close to zero for left-out observations. The coverage of 95\% and 80\% prediction intervals are as expected and symmetrical. 

Table~\ref{tb-val-res-est} shows results for the comparison between model estimates obtained based on the full dataset and based on the training set for the out-of-sample validation exercises in (i) M1: for the SRB $\Theta_{c,t}$, and (ii) M2: for the SRB $\Theta_{c,t}$ and the inflation $\delta_c\Omega_{c,t}$. Median errors and the median absolute errors are close to zero. The proportions of updated estimates that fall above or below their respective credible intervals constructed based on the training set are reasonable and mostly within the expected range, with at most two countries' estimates falling outside their respective bounds.

\begin{table}[htpb]
\caption[Validation results for left-out observations]{\textbf{Validation results for left-out observations.} Error is defined as the difference between a left-out observation and the posterior median of its predictive distribution. M1: model without inflation term (Section~\ref{sec_method_step2}). M2: model with inflation term (Section~\ref{sec_method_step3}). Prediction from 1970: predicting sex ratio transition based on prediction setup (Section~\ref{sec_method_predict}).} 
\label{tb-val-res-ppd}
\centering
\begin{tabular}{|l| cc | c | c |} \hline 
\bf{Left-out observations}& \multicolumn{2}{c |}{\bf{M1}} & \bf{M2} & \bf{M2} \\ 
$\boldsymbol{y_j}$& \textbf{Recent obs.} & \textbf{Random}				& \textbf{Recent obs.} & \textbf{Prediction from 1970}\\ \hline
\# Country in training dataset & 176 & 184 & 29 & 29 \\ 
\# Country in test dataset 	 	& 143 & 169 & 28 & 29 \\ \hline
 Median error 						& 0.000 & -0.001 & -0.003 & 0.010\\ 
 Median absolute error 			 	& 0.015 & 0.012 & 0.020 & 0.027\\ 
\hline
 Below 95\% prediction interval (\%) & 2.7 & 3.0 & 4.6 & 1.4\\ 
 Above 95\% prediction interval (\%) & 3.6 & 2.3 & 1.7 & 3.6\\ 
 \bf{Expected (\%)} & \bf{2.5} & \bf{2.5} & \bf{2.5} & \bf{2.5} \\
 \hline
 Below 80\% prediction interval (\%) & 9.7 & 9.8 & 11.3 & 7.0\\ 
 Above 80\% prediction interval (\%) & 10.2 & 8.2 & 8.6 & 14.0\\ 
 \bf{Expected (\%)} & \bf{10} & \bf{10} & \bf{10} & \bf{10} \\
 \hline
\end{tabular}
\end{table}

\begin{table}[htpb]
\caption[Validation results for $\Theta_{c,t}$ estimates based on training set for M1, and for $\Theta_{c,t}$ and $\Omega_{c,t}\delta_c$ estimates based on training set for M2]{\textbf{Validation results for $\boldsymbol{\Theta_{c,t}}$ estimates based on training set for M1, and for $\Theta_{c,t}$ and $\Omega_{c,t}\delta_c$ estimates based on training set for M2.} Error is defined as the differences between an estimate based on full dataset and training set. The percentages (\%) of countries in which the SRB median estimates based on the full dataset fall below or above their respective 95\% and 80\% credible intervals based on the training set are reported. Numbers in the parentheses indicate the number of countries with median estimates based on the full dataset that fall below or above their respective 95\% and 80\% credible intervals based on the training set. M1: model without inflation (Section~\ref{sec_method_step2}). M2: model with inflation (Section~\ref{sec_method_step3}).} 
\label{tb-val-res-est}
\centering
\begin{tabular}{|l|ccc|ccc|ccc|} \hline
\bf{Validation (Out-of-Sample)} & \multicolumn{3}{c|}{\bf{M1}}& \multicolumn{6}{c|}{\textbf{M2}}\\ \hline
\# Country in training dataset & \multicolumn{3}{c|}{176} & \multicolumn{6}{c|}{29}\\
\# Country in full dataset & \multicolumn{3}{c|}{202} & \multicolumn{6}{c|}{29}\\\hline
Outcome of interest & \multicolumn{3}{c|}{$\boldsymbol{\Theta_{c,t}}$}& \multicolumn{3}{c|}{$\boldsymbol{\Theta_{c,t}}$} & \multicolumn{3}{c|}{$\boldsymbol{\delta_c\Omega_{c,t}}$}\\ 
Reference year & \bf{1995} & \bf{2005} & \bf{2015} & \bf{1995} & \bf{2005} & \bf{2015} & \bf{1995} & \bf{2005} & \bf{2015} \\\hline
 Median error 			& 0.001 & 0.000 & 0.001 & 0.000 & 0.000 & -0.002 & 0.000 & 0.000 & 0.000 \\ 
 Median absolute error 	& 0.003 & 0.004 & 0.004& 0.001 & 0.000 & 0.003 & 0.000 & 0.000 & 0.000 \\ 
\hline
 Below 95\% credible interval (\%) & 2.4 (5) & 2.4 (5) & 0.9 (2) & 0.0 & 0.0 & 3.4 (1) & 0.0 & 0.0 & 3.4 (1)\\ 
 Above 95\% credible interval (\%) & 2.4 (5) & 2.8 (6) & 2.8 (6) & 0.0 & 0.0 & 0.0 & 0.0 & 0.0 & 0.0\\ 
 \bf{Expected proportions (\%)} & $\leq$\bf{2.5}& $\leq$\bf{2.5} & $\leq$\bf{2.5}& $\leq$\bf{2.5}& $\leq$\bf{2.5} & $\leq$\bf{2.5}& $\leq$\bf{2.5}& $\leq$\bf{2.5} & $\leq$\bf{2.5}\\
 \hline
 Below 80\% credible interval (\%) & 9.0 (19) & 9.0 (19) & 4.7 (10) & 0.0 & 0.0 & 6.9 (2) & 0.0 & 0.0 & 3.4 (1)\\ 
 Above 80\% credible interval (\%) & 8.0 (15) & 9.4 (20) & 8.0 (17) & 0.0 & 3.4 (1) & 3.4 (1) & 0.0 & 0.0 & 3.4 (1)\\ 
 \bf{Expected proportions (\%)} & $\leq$\bf{10}& $\leq$\bf{10}& $\leq$\bf{10}& $\leq$\bf{10}& $\leq$\bf{10}& $\leq$\bf{10} & $\leq$\bf{10}& $\leq$\bf{10}& $\leq$\bf{10}\\
 \hline
\end{tabular}
\end{table}

\clearpage
\section{Discussion}\label{sec_discussion}
We described a Bayesian hierarchical time series model for producing SRB estimates and scenario-based projections for all countries from 1950 to 2100. The model produces estimates of SRB baselines by country and region, capturing variations by ethnicity. The model also captures SRB variations under normal circumstance and in settings with sex-selective abortion. The SRB inflation model includes a country-specific indicator to estimate the probability of SRB inflation and identifies countries with strong statistical evidence of SRB inflation in a fully Bayesian approach. Furthermore, the inflation model provides a reproducible way to project SRB inflation based on three scenarios that are data-driven and model-based. Model validation exercises suggest that the SRB models are reasonably well calibrated and have satisfactory predictive performance.

We provide scenario-based projections that quantify the SRB inflation if high fertility countries with a son preference would also experience sex ratio transitions, similar to the ones observed so far. In the 17 countries at risk of sex ratio imbalance but without strong statistical evidence of SRB inflation, scenario-based projections S2 and S3 indicate that future sex ratio transitions may happen later this century, when fertility in at-risk countries reaches low levels.

Our SRB estimation model is based on several model assumptions and is subject to limitations. We summarize the main ones as discussed in detail in our prior study \cite{chao2019systematic} here. Firstly, the SRB baselines are modeled according to regional groupings and do not depend on external indicators. Secondly, aside from the fertility squeeze effect, we are not able to incorporate additional factors in the SRB inflation model that may affect SRB imbalance. Thirdly, out of the 212 countries included in the study, some do not have information to indicate whether they are at risk of SRB imbalance (contributing  3.2\% of the global births in 1970--2017). We assumed that these countries have no risk of SRB inflation which is a limitation for monitoring in those specific countries. Lastly, we select countries with risk of SRB inflation prior to model fitting, as opposed to incorporating the selection in the model. This approach was motivated by the need to model the natural fluctuations $\eta_{c,t}$ and SRB inflation $\delta_c\Omega_{c,t}$ sequentially to avoid identifiability issues. 

The SRB projection scenarios are subject to additional limitations. Firstly, given variability across countries that have started their transition, there is substantial uncertainty associated with the inflation predictions for countries prior to observing country-specific transition data. Secondly, given that the SRB inflation has been (partially) observed in 12 countries only, the evidence base to extrapolate to other countries is limited. Hence long-term projections up to 2100 are subject to additional uncertainty that is not captured in our approach. For example, given that no country so far has experienced a re-occurrence of SRB inflation, we do not consider re-occurrence of a sex ratio transition. Lastly, the calculation of missing female births is based on the UN WPP projections of the number of total births, as opposed to projections of births that take account of lower or higher SRBs resulting in larger or smaller female cohorts, and subsequently, larger or smaller number of births.

This study focuses on modeling national-level SRB estimation and projection. However, insights from national-level analyses may not be sufficient in national populations with great demographic heterogeneity. Several studies have shown that the SRB can differ greatly across geographic locations or other subpopulations within a country \cite{chao2019levels,guilmoto2015mapping,jiang2019change}. For these countries, SRB imbalances may start to decline in some regions while increasing in others, i.e. in higher fertility regions.  Projections on the national level may mask variability in the sex ratio imbalance at birth across sub-regions within countries. Constructing subnational scenario-based projections in such settings will provide additional important insights into future missing female births.

While the assumptions made for the scenario-based projections presented in this study may be hypothetical, the associated projections of missing girls are important illustrations of the potential burden of future prenatal sex discrimination and the need to monitor SRBs in countries with son preference. The Sustainable Development Goals (see \url{http://www.un.org/sustainabledevelopment/sustainable-development-goals/}) include the goal to ``achieve gender equality and empower all women and girls'' by 2030. Monitoring and projecting the sex ratio at birth is an essential part in protecting the gender equality at the prenatal stage. Our scenario-based projections underscore the importance of the monitoring of the sex ratio at birth over time, especially in countries with ongoing inflations and countries where future sex imbalances may occur, to avoid future aborting of girls in favor of male offspring.

\begin{appendix}
\section{Model specification and priors}\label{app_model}
We fit different variations of the general SRB process model equation
\begin{eqnarray*}
\Theta_{c,t} &=& \beta_c \eta_{c,t}(\boldsymbol{\phi}) + \delta_c \Omega_{c,t}(\boldsymbol{\zeta}),
\end{eqnarray*}
with details provided in the remainder of this section.

[M1] Model stage 1: the model for the SRB in country-years without SRB inflation:
\begin{eqnarray*}
\Theta_{c,t} &=& \beta_c \eta_{c,t}(\boldsymbol{\phi}), \text{ for } c \in \mathcal{C}^{\text{base}}, \text{ for }t \in \{1950, \cdots, 2100\},\\
\log(\beta_c) | \beta_{r[c]}^{(\text{region})}, \sigma_\beta &\sim& \mathcal{N}\left(\log\left(\beta_{r[c]}^{(\text{region})}\right), \sigma_\beta^2\right), \text{ for } c \in \mathcal{C},\\
\beta_r^{(\text{region})} &\overset{\text{i.i.d.}}\sim& \mathcal{U}(1, 1.1), \text{ for } r \in \mathcal{R},\\
\log(\eta_{c,t}) | \boldsymbol{\phi} &\sim& \mathcal{N}(0, (1-\rho^2)/ \sigma_\epsilon^2), \text{ for }t=1950,\\
\log(\eta_{c,t}) &=& \rho\log(\eta_{c,t-1}) + \epsilon_{c,t}, \text{ for }t\in\{1951, \cdots, 2100\},\\
\epsilon_{c,t} | \sigma_\epsilon &\overset{\text{i.i.d.}}\sim& \mathcal{N}(0, \sigma_\epsilon^2),\\
\sigma_\beta &\sim& \mathcal{U}(0, 0.05),\\
\rho &\sim& \mathcal{U}(0, 1),\\
\sigma_\epsilon &\sim& \mathcal{U}(0, 0.05).
\end{eqnarray*}
where $\mathcal{C}^{\text{base}}$ refers to the set of countries not identified with risk of SRB inflation and $\mathcal{C}$ is the set of all the 212 countries. $\mathcal{R}$ is the set of all the 10 regions. $\boldsymbol{\phi}=\{ \rho, \sigma_{\epsilon}\}$ is the vector of hyper parameters related to $\eta_{c,t}$. $\mathcal{U}(a, b)$ refers to a continuous uniform distribution with lower and upper bounds at $a$ and $b$ respectively. 

[M2] Model stage 2: the SRB model for country-years with potential SRB inflation is:
\begin{eqnarray*}
\Theta_{c,t} &=& \hat{\beta}^{\text{\tiny [M1]}}_c \eta_{c,t}\left(\boldsymbol{\hat{\phi}}^{\text{\tiny [M1]}} \right) + \delta_c \Omega_{c,t}(\boldsymbol{\zeta}), \text{ for } c \in \mathcal{C}^{\text{inflation}}\cup \mathcal{C}^{\text{future-inf}}, \text{ for }t \in \{1950, \cdots, 2100\},\\
\Omega_{c,t} &=&
\left\{\begin{matrix}
\xi_c (t - \gamma_{0,c}) / \lambda_{1,c}, & \gamma_{0,c} < t < \gamma_{1,c} \\ 
 \xi_c, & \gamma_{1,c} <t < \gamma_{2,c} \\
 \xi_c - \xi_c (t -\gamma_{2,c}) / \lambda_{3,c}, & \gamma_{2,c} < t < \gamma_{3,c} \\
0, & t < \gamma_{0,c} \text{ or } t > \gamma_{3,c}
\end{matrix}\right.,\text{ where}\\
\gamma_{1,c} &=& \gamma_{0,c} + \lambda_{1,c},\\
\gamma_{2,c} &=& \gamma_{1,c} + \lambda_{2,c},\\
\gamma_{3,c} &=& \gamma_{2,c} + \lambda_{3,c},\\
\xi_c | \mu_{\xi}, \sigma_{\xi} &\sim& \mathcal{N}(\mu_{\xi}, \sigma_{\xi}^2)T(0, ), \text{ for } c \in \mathcal{C}^{\text{inflation}}\cup \mathcal{C}^{\text{future-inf}},\\
\lambda_{1,c} | \mu_{\lambda1}, \sigma_{\lambda1} &\sim& \mathcal{N}(\mu_{\lambda1}, \sigma_{\lambda1}^2)T(0, ), \text{ for } c \in \mathcal{C}^{\text{inflation}}\cup \mathcal{C}^{\text{future-inf}},\\
\lambda_{2,c} | \mu_{\lambda2}, \sigma_{\lambda2}&\sim& \mathcal{N}(\mu_{\lambda2}, \sigma_{\lambda2}^2)T(0, ), \text{ for } c \in \mathcal{C}^{\text{inflation}}\cup \mathcal{C}^{\text{future-inf}},\\
\lambda_{3,c} | \mu_{\lambda3}, \sigma_{\lambda3} &\sim& \mathcal{N}(\mu_{\lambda3}, \sigma_{\lambda3}^2)T(0, ), \text{ for } c \in \mathcal{C}^{\text{inflation}}\cup \mathcal{C}^{\text{future-inf}},\\
\gamma_{0,c}  | \sigma_{\gamma} &\sim& t_3(x_c, \sigma_{\gamma}^2, \nu=3)T(z_c, ), \text{ for } c \in \mathcal{C}^{\text{inflation}}\cup \mathcal{C}^{\text{future-inf}},\\
\delta_c | \pi_c &\sim& \mathcal{B}(\pi_c), \text{ for } c \in \mathcal{C}^{\text{inflation}}\cup \mathcal{C}^{\text{future-inf}},\\
\text{logit}(\pi_c) | \mu_\pi, \sigma_\pi &\sim& \mathcal{N}(\mu_\pi, \sigma_\pi^2), \text{ for } c \in \mathcal{C}^{\text{inflation}}\cup \mathcal{C}^{\text{future-inf}},\\
\mu_{\xi} &\sim& \mathcal{U}(0, 2), \\
\mu_{\lambda1} &\sim& \mathcal{U}(0, 40),\\
\mu_{\lambda2} &\sim& \mathcal{U}(0, 40),\\
\mu_{\lambda3} &\sim& \mathcal{U}(0, 40),\\
\text{inverse-logit}(\mu_\pi) &\sim& \mathcal{U}(0, 1),\\
\sigma_{\lambda1} &\sim& \mathcal{U}(1, 10),\\
\sigma_{\lambda2} &\sim& \mathcal{U}(1, 10),\\
\sigma_{\lambda3} &\sim& \mathcal{U}(1, 10),\\
\sigma_{\xi} &\sim& \mathcal{U}(0, 2),\\
\sigma_{\gamma} &\sim& \mathcal{U}(0, 10),\\
\sigma_\pi &\sim& \mathcal{U}(0, 2).
\end{eqnarray*}
$\hat{\beta}^{\text{\tiny [M1]}}_c$ is the posterior median for the national baseline from M1 and $\boldsymbol{\hat{\phi}}^{\text{\tiny [M1]}}=\{ \hat{\rho}^{\text{\tiny [M1]}}, \hat{\sigma_{\epsilon}}^{\text{\tiny [M1]}} \}$ the vector of posterior medians of $\boldsymbol{\phi}$. $\boldsymbol{{\zeta}} = \{\mu_{\xi}, \sigma_{\xi}, \mu_{\lambda1}, \sigma_{\lambda1}, \mu_{\lambda2}, \sigma_{\lambda2}, \mu_{\lambda3}, \sigma_{\lambda3}, \sigma_{\gamma}\}$ is the vector of hyper parameters related to $\Omega_{c,t}$. $\mathcal{N}(\cdot)T(a, )$ refers to a truncated normal distribution with lower truncation at $a$.

The data quality model for SRB observations is:
\begin{eqnarray*}
\log(y_i)|\Theta_{c[i], t[i]},\omega_{s[i]} &\sim& \mathcal{N}\left( \log(\Theta_{c[i], t[i]}), \omega_{s[i]}^2 + v_i^2\right), \text{ for }i \in \{ 1, \cdots, n \},\\
\omega_s &=& 0, \text{ for }s=\text{CRVS/SRS},\\
\omega_s &\sim& \mathcal{U}(0,0.5), \text{ for }s \in \{\text{Census, DHS, Other DHS, Other}\}
\end{eqnarray*}

\paragraph*{Scenario-based projections}
The $g$-th scenario-based projection trajectory for country $c$ and year $t \in \{2018, \hdots, 2100\}$, $^{(g)}{\Theta}^{\text{\tiny [S]}}_{c,t}$, is obtained as follows:
\begin{equation*}
^{(g)}{\Theta}^{\text{\tiny [S]}}_{c,t} =
\begin{cases}
^{(g)}{\beta}^{\text{\tiny [M1]}}_c \cdot ^{(g)}{\eta}^{\text{\tiny [M1]}}_{c,t} & \text{ for } c \in \mathcal{C}^{\text{base}}, \text{ S} \in \{\text{S1}, \text{S2}, \text{S3}\},\\ 
^{(g)}{\beta}^{\text{\tiny [M1]}}_c
\cdot ^{(g)}{\eta}^{\text{\tiny [M2]}}_{c,t} +
^{(g)}{\delta}^{\text{\tiny [M2]}}_c \cdot 
^{(g)}{\Omega}^{\text{\tiny [M2]}}_{c,t} & \text{ for } c \in \mathcal{C}^{\text{inflation}}, \text{ S} \in \{\text{S1}, \text{S2}, \text{S3}\},\\ 
^{(g)}{\beta}^{\text{\tiny [M1]}}_c \cdot ^{(g)}{\eta}^{\text{\tiny [M3]}}_{c,t} & \text{ for } c \in \mathcal{C}^{\text{future-inf}}, \text{ S}=\text{S1},\\ 
^{(g)}{\beta}^{\text{\tiny [M1]}}_c
\cdot ^{(g)}{\eta}^{\text{\tiny [M2]}}_{c,t} +
^{(g)}{\delta}^{\text{\tiny [M2]}}_c \cdot 
^{(g)}{\Omega}^{\text{\tiny [M2]}}_{c,t} & \text{ for } c \in \mathcal{C}^{\text{future-inf}}, \text{ S}=\text{S2},\\ 
^{(g)}{\beta}^{\text{\tiny [M1]}}_c \cdot ^{(g)}{\eta}^{\text{\tiny [M4]}}_{c,t} + ^{(g)}{\Omega}^{\text{\tiny [M4]}}_{c,t} & \text{ for } c \in \mathcal{C}^{\text{future-inf}}, \text{ S}=\text{S3},
\end{cases}    
\end{equation*}
where $\mathcal{C}^{\text{base}}$ refers to countries without risk of SRB inflation, $\mathcal{C}^{\text{inflation}}$ to countries with strong statistical evidence of SRB inflation, and $\mathcal{C}^{\text{future-inf}}$ to countries at risk of SRB inflation but without strong statistical evidence of SRB inflation.

The models used are defined as follows:
\begin{itemize}
\item M1: Model without inflation term, $\Theta_{c,t} = \beta_c \eta_{c,t}(\boldsymbol{\phi})$, fit to dataset $\boldsymbol{y}^{(\text{risk-free})}$, as described in Section~\ref{sec_method_step2};
\item M2: Model with inflation term, $\Theta_{c,t} = \hat{\beta}^{\text{\tiny [M1]}}_c \eta_{c,t}\left(\boldsymbol{\hat{\phi}}^{\text{\tiny [M1]}} \right) + \delta_c \Omega_{c,t}(\boldsymbol{\zeta})$, fit to all data from at-risk countries $\boldsymbol{y}^{(\text{at-risk})}$, as described in Section~\ref{sec_method_step3};
\item M3: Model without inflation term, $\Theta_{c,t} = \hat{\beta}^{\text{\tiny [M1]}}_c \eta_{c,t}\left(\boldsymbol{\hat{\phi}}^{\text{\tiny [M1]}} \right)$, fit to country-specific dataset $\boldsymbol{y}^{(c)}$, used for scenario S1 for $c \in \mathcal{C}^{\text{future-inf}}$;
\item M4: Model with inflation term, $\Theta_{c,t} = \hat{\beta}^{\text{\tiny [M1]}}_c \eta_{c,t}\left(\boldsymbol{\hat{\phi}}^{\text{\tiny [M1]}} \right) + \Omega_{c,t}\left(\boldsymbol{\hat{\zeta}}^{\text{\tiny [M2]}} \right)$, fit to country-specific dataset $\boldsymbol{y}^{(c)}$, used for scenario S3 for $c \in \mathcal{C}^{\text{future-inf}}$.
\end{itemize}

\section{Computation}\label{app_compute}
We obtain posterior samples using a Markov chain Monte Carlo (MCMC) algorithm, implemented in the open source software \texttt{R 3$\cdot$6$\cdot$1} \cite{R2019} and \texttt{JAGS 4$\cdot$0$\cdot$1} \cite{plummer2003}, using \texttt{R}-packages \texttt{rjags} \cite{rjags}, \texttt{R2jags} \cite{r2jags} and \texttt{MCMCpack} \cite{r_MCMCpack}. Convergence of the MCMC algorithm and the sufficiency of the number of samples obtained are checked through visual inspection of trace plots and convergence diagnostics of Gelman and Rubin \cite{gelmanrubin1992}, implemented in the \texttt{coda} \texttt{R}-package \cite{coda}. Table~\ref{tab-computing-MCMC} summarizes the MCMC specifications for model runs.

\begin{figure}[htbp]
\centering
\includegraphics[width=0.9\linewidth]{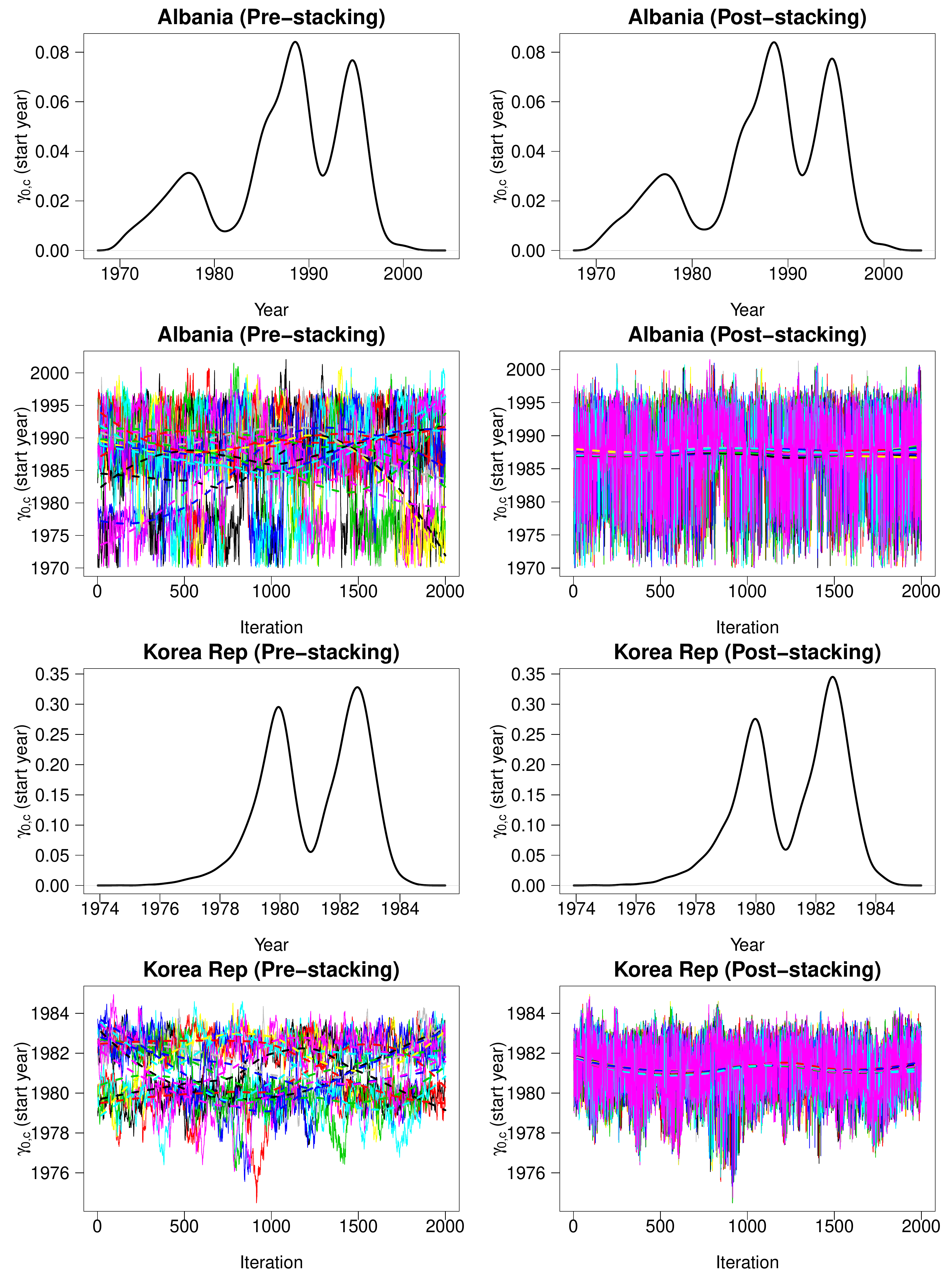}
\caption[Posterior density and trace plots of start years pre- and post-stacking, Albania and Republic of Korea]{\textbf{Posterior density and trace plots of start years pre- and post-stacking, Albania and Republic of Korea.}}
\label{fig_stacking}
\end{figure}

Due to the multimodal nature of the posterior distributions for start year parameters in Albania and Republic of Korea, we apply a stacking approach as discussed in \cite{yao2020stacking} to obtain representative samples from the posterior distribution. Figure~\ref{fig_stacking} shows the posterior density and trace plots of start years pre- and post-stacking for Albania and Republic of Korea.

In the stacking approach, a weight $w_k$ is assigned to each parallel MCMC chain indexed by $k \in \{1,2,\hdots, K\}$. The chain-specific weights $w_k$ are optimized to maximize the leave-one-out cross validation performance of the distribution formed from the weighted average of the sets of simulation draws \cite{yao2020stacking}. The weight for each chain is calculated by solving the following:
\begin{equation*}
\hat{w} = \text{arg }\underset{w \in \mathbb{S}(N)}{\text{max}} \sum_{i=1}^n \log \sum_{k=1}^K w_k f_k(y_i | y_{-i}),% + \log f_{\text{prior}}(w).%\left( r_i - \sum_{k=1}^K w_k \hat{f}_k^{(-i)}(r_i) \right)^2.
\end{equation*}
where $\mathbb{S}(N) = \{ w: 0 \leq w_k \leq 1, \forall 1 \leq k \leq N; \sum_{k=1}^K w_k = 1\}$ and $f_k(y_i | y_{-i})$ refers to the pointwise leave-one-out (loo) log predictive density for $y_i$ based on the posterior samples of model parameters in chain $k$. We use the \texttt{R}-package \texttt{loo} \cite{vehtari2018loo} to approximate $f_k(y_i | y_{-i})$. After obtaining the weights, we re-sample the parallel chains based on their respective stacking weight. The greater the stacking weight for a chain, the higher the probability that the posterior samples from the chain are re-sampled. The resulting post-stacking samples are used to produce estimates and credible bounds.

\begin{table}
	\caption[MCMC specifications]{MCMC specifications.}
	\label{tab-computing-MCMC}
\centering
\begin{tabular}{| l | c | c | c | c | c |}
	\hline
	\bf{MCMC Specifications} & \multicolumn{3}{c|}{\bf{Normal Model}} & \multicolumn{2}{c|}{\bf{Inflation Model}}\\ 
				& \textbf{Full} 	& \multicolumn{2}{c|}{\textbf{Validation}}& \textbf{Full} 	& \textbf{Validation} \\ 
				&		& \textbf{Out-of-Sample} & \textbf{In-Sample}		 &			& \textbf{Out-of-Sample}\\ \hline
	\# Chains 	& 8		& 8 			& 8				 & 14 		& 8 \\ \hline
	\# Burn-in 	& 8,000 & 8,000 		& 8,000			 & 7,600 	& 1,000 \\ \hline
	\# Thinning & 20 	& 20 			& 20			 & 10 		& 2 \\ \hline
	\# Posterior samples per parameter & 4,000 & 4,000 & 4,000 & 28,000 & 11,000 \\ \hline
\end{tabular}
\end{table}

\section{Model comparison for M2 related to estimating national baselines sequentially or jointly}\label{app_model_jointseq}

The model equation for M2 (Section~\ref{sec_method_step3}) is as follows: 
\begin{eqnarray*}
\Theta_{c,t} &=& \hat{\beta}^{\text{\tiny [M1]}}_c \eta_{c,t}\left(\boldsymbol{\hat{\phi}}^{\text{\tiny [M1]}} \right) + \delta_c \Omega_{c,t}(\boldsymbol{\zeta}),
\end{eqnarray*}
where $\hat{\beta}^{\text{\tiny [M1]}}_c$ refers to the point estimate for the national baseline from model M1. We compare the result from M2 to an alternative model, referred to as M2Joint, in which $\beta_c$ and $\Omega_{c,t}$ are estimated jointly. Specifically, in model M2Joint, we use country-specific priors on $\beta_c$:
\begin{eqnarray*}
\Theta_{c,t} &=& \beta_c \eta_{c,t}\left(\boldsymbol{\hat{\phi}}^{\text{\tiny [M1]}} \right) + \delta_c \Omega_{c,t}(\boldsymbol{\zeta}),\\
\beta_c &\sim& \mathcal{N}\left( \hat{\beta}_c^{\text{(region) \tiny [M1]}}, \left(\hat{\sigma}^{\text{\tiny [M1]}}_{\beta}\right)^2 \right),
\end{eqnarray*}
where $\hat{\beta}_c^{\text{(region) \tiny [M1]}}$ and $\hat{\sigma}^{\text{\tiny [M1]}}_{\beta}$ are posterior medians obtained from M1. 

\begin{figure}[htbp]
\centering
\begin{tabular}{c}
\includegraphics[width=0.8\linewidth,page=6]{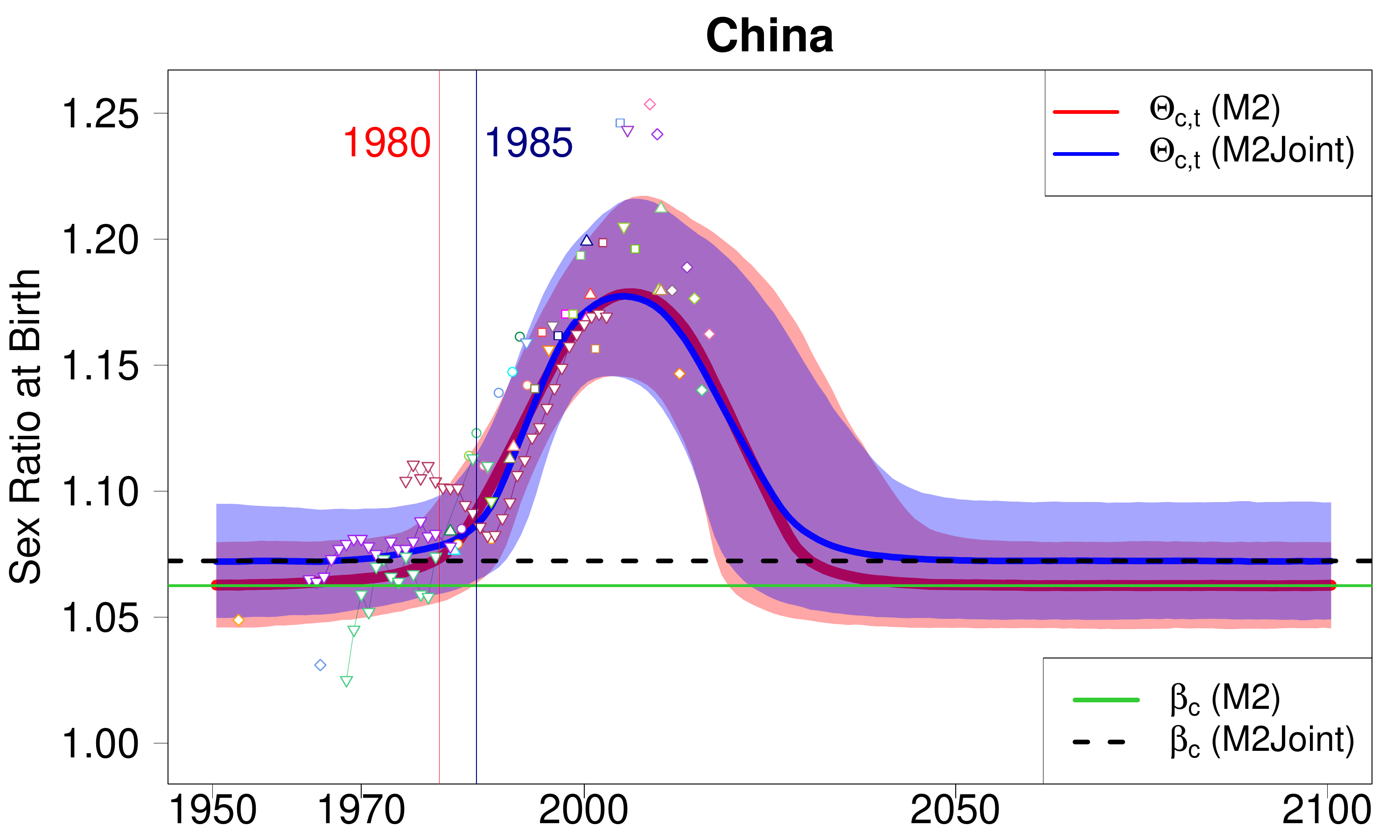}\\
\includegraphics[width=0.8\linewidth,page=11]{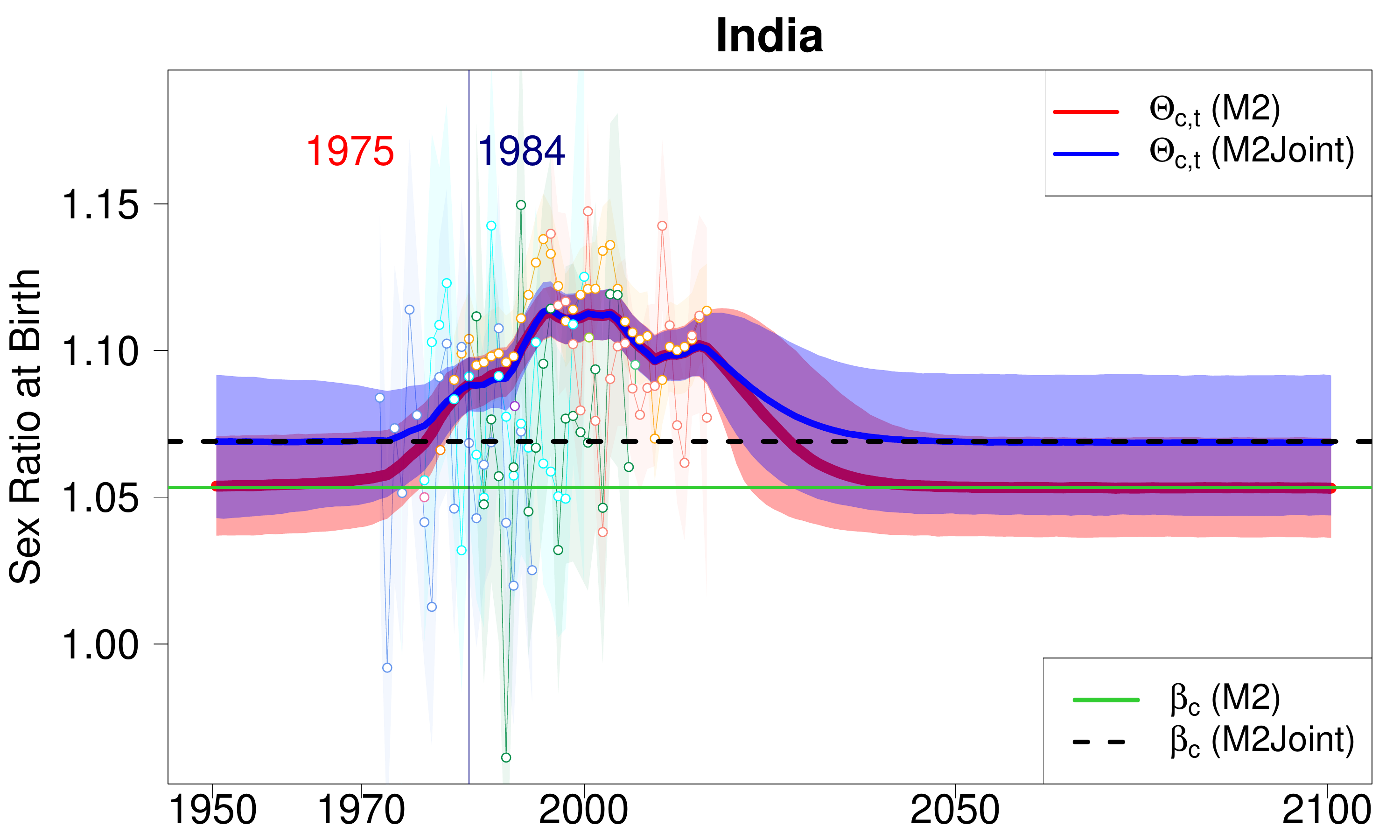}
\end{tabular}
\caption[SRB results comparison between M2 and M2Joint for China and India]{\textbf{SRB results comparison between M2 and M2Joint for China and India.} M2: estimating $\beta_c$ and $\Omega_{c,t}$ sequentially (model presented in Section~\ref{sec_method_step3}; in red). M2Joint: estimating $\beta_c$ and $\Omega_{c,t}$ jointly (in blue). SRB median estimates/projections and 95\% credible intervals (curves and shades) from M2 (in red) and from M2Joint (in blue). Median estimates of $\beta_c$ from M2 (light green line). $\beta_c$ model results for M2Joint (black dashed line). Median estimates of the start year of inflation from M2 (red vertical lines) and from M2Joint (blue vertical lines), with years stated. SRB observations are displayed with dots.}
\label{fig_fullbaye}
\end{figure}

M2Joint model fittings result in estimates of start years $\gamma_{0,c}$ in China and India that are deemed too late based on external information regarding the timing of first occurrences of sex-selective abortion.
Figure~\ref{fig_fullbaye} compares results between M2 and M2Joint for both countries. There are noticeable difference in the $\beta_c$ between the two models. The estimated $\beta_c$'s from model M2Joint are higher than those from model M2. Consequently, the estimated start years based on model M2Joint are later than the estimated start years based on model M2. For China, the M2-based start year is 1980 compared to 1985 for M2Joint. For India, the M2-based start year is 1975 compared to 1984 for M2Joint. In China, the ``one-child policy'' was implemented in 1980 and marked the start of a nationwide abortion campaign. Given the timing of this campaign, an estimated start year of 1985 for SRB inflation based on model M2Joint is deemed to be biased upwards for China. In India, prenatal diagnosis (PD) became available soon after abortion was legalized in 1971. PD was introduced in India as a method for detecting fetal abnormalities but was soon used for prenatal sex selection \cite{allahbadia200250,tandon2006female}. Since then, the combination of prenatal sex determination and selective abortion has been widely used for the systematic elimination of females fetuses \cite{madan2013impact}. Given the timing of the introduction of PD in 1971, start year 1984 based on model M2Joint is deemed biased upwards.

\end{appendix}

\section*{Acknowledgements}
The authors would like to thank the anonymous referees, an Associate Editor and the Editor for their insightful and constructive comments that improved the quality of this paper. This work was supported by a research grant from the National University of Singapore. The study described is solely the responsibility of the authors and does not necessarily represent the official views of the United Nations.

\clearpage
\section{Supplement materials}
\subsection{Supplement figure A: Scenario-based SRB projection during 1950--2100, by country}
Page 31--41.

SRB median estimates/projections (curve) and 95\% credible intervals (shades) for Scenario 1 $\Theta_{c,t}^{\text{\tiny [S1]}}$ (in red), Scenario 2 $\Theta_{c,t}^{\text{\tiny [S2]}}$ (in blue) and Scenario 3 $\Theta_{c,t}^{\text{\tiny [S3]}}$ (in green), and median estimates of the national baselines $\beta_c$ (black dashed horizontal line). The median estimates and projections of total fertility rate (TFR) from the UN WPP 2019 are added (blue squared dots). The median estimates of inflation start year $\gamma_{0,c}$ and end year $\gamma_{3,c}$ are the vertical lines. The TFR value in the year $\gamma_{0,c}$ is shown.

\subsection{Supplement figure B: Scenario 1 SRB projection during 1950--2100 for all countries}
Page 42--113.

SRB median estimates/projections and 95\% credible intervals for Scenario 1 $\Theta_{c,t}^{\text{\tiny [S1]}}$ (red curve and shades), median estimates of the regional baselines $\beta_{r[c]}^{(\text{region})}$ (dark green horizontal line), median estimates of the national baselines $\beta_c$ (light green horizontal line). SRB observations are displayed with dots and observations are connected with lines when obtained from the same source. Shaded areas around observation series represent the sampling variability in the series (quantified by two times the stochastic/sampling standard errors). Model results are shown before 1950 if observations are available prior 1950.

\clearpage
\bibliography{srb}
\clearpage

\begin{figure}[htbp]
\begin{centering}
\caption[Scenario-based SRB projection during 1950--2100, by country]{\textbf{Scenario-based SRB projection during 1950--2100, by country.} SRB median estimates/projections (curve) and 95\% credible intervals (shades) for Scenario 1 $\Theta_{c,t}^{\text{\tiny [S1]}}$ (in red), Scenario 2 $\Theta_{c,t}^{\text{\tiny [S2]}}$ (in blue) and Scenario 3 $\Theta_{c,t}^{\text{\tiny [S3]}}$ (in green), and median estimates of the national baselines $\beta_c$ (black dashed horizontal line). The median estimates and projections of total fertility rate (TFR) from the UN WPP 2019 are added (blue squared dots). The median estimates of inflation start year $\gamma_{0,c}$ and end year $\gamma_{3,c}$ are the vertical lines. The TFR value in the year $\gamma_{0,c}$ is shown.}
\label{sup_fig_scenario_proj}
\end{centering}
\end{figure}

\afterpage{
\clearpage
\centering
\includepdf[scale=0.78,pages = -, pagecommand={\thispagestyle{plain}}]{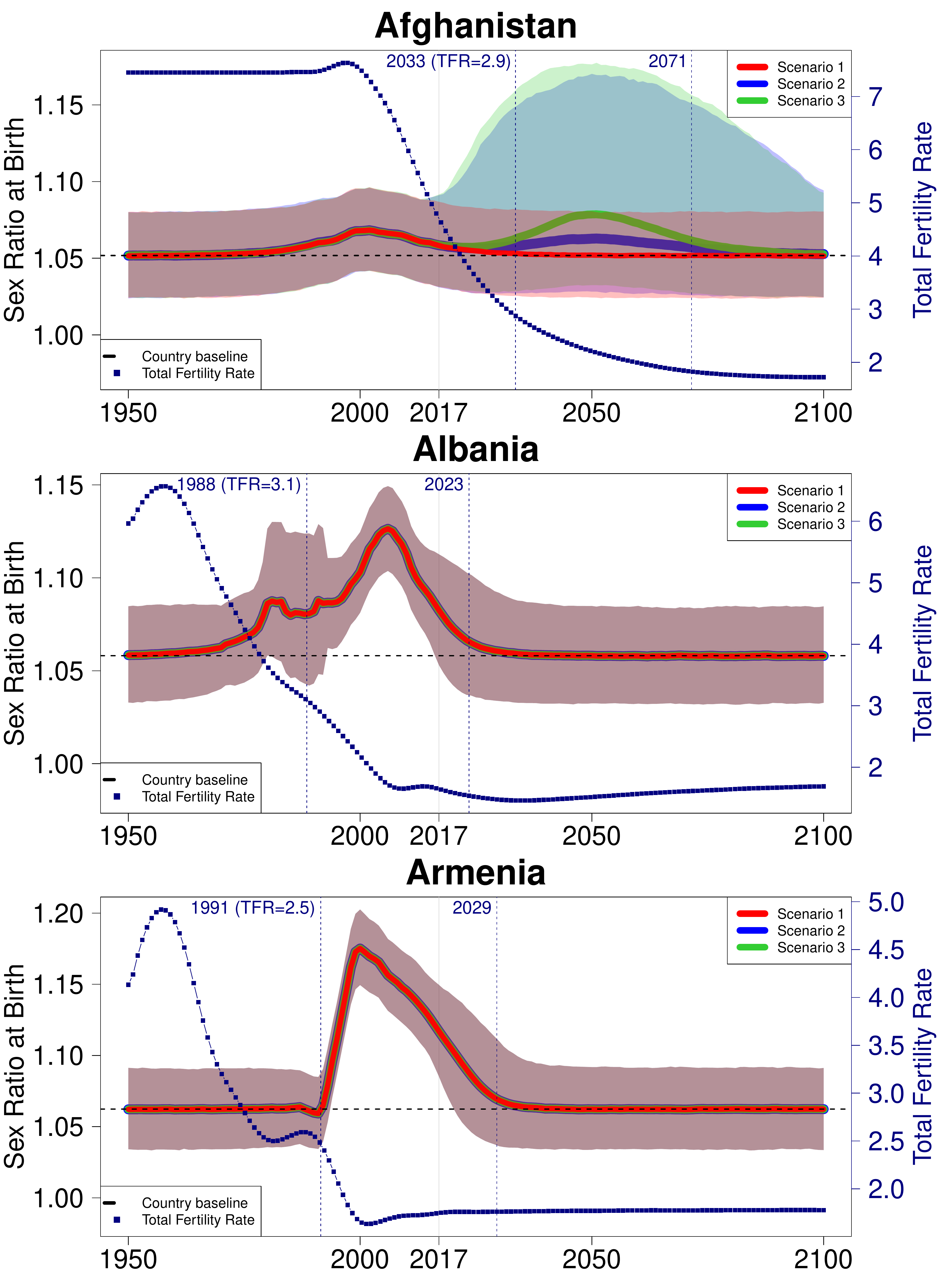}
}

\afterpage{
\clearpage
\centering
\begin{figure}[htbp]
\begin{centering}
\caption[Scenario 1 SRB projection during 1950--2100 for all countries]{\textbf{Scenario 1 SRB projection during 1950--2100 for all countries.} SRB median estimates/projections and 95\% credible intervals for Scenario 1 $\Theta_{c,t}^{\text{\tiny [S1]}}$ (red curve and shades), median estimates of the regional baselines $\beta_{r[c]}^{(\text{region})}$ (dark green horizontal line), median estimates of the national baselines $\beta_c$ (light green horizontal line). SRB observations are displayed with dots and observations are connected with lines when obtained from the same source. Shaded areas around observation series represent the sampling variability in the series (quantified by two times the stochastic/sampling standard errors). Model results are shown before 1950 if observations are available prior 1950.}
\label{fig_allcountry_s1}
\end{centering}
\end{figure}

}

\afterpage{
\clearpage
\centering
\includepdf[scale=0.78,pages = -, pagecommand={\thispagestyle{plain}}]{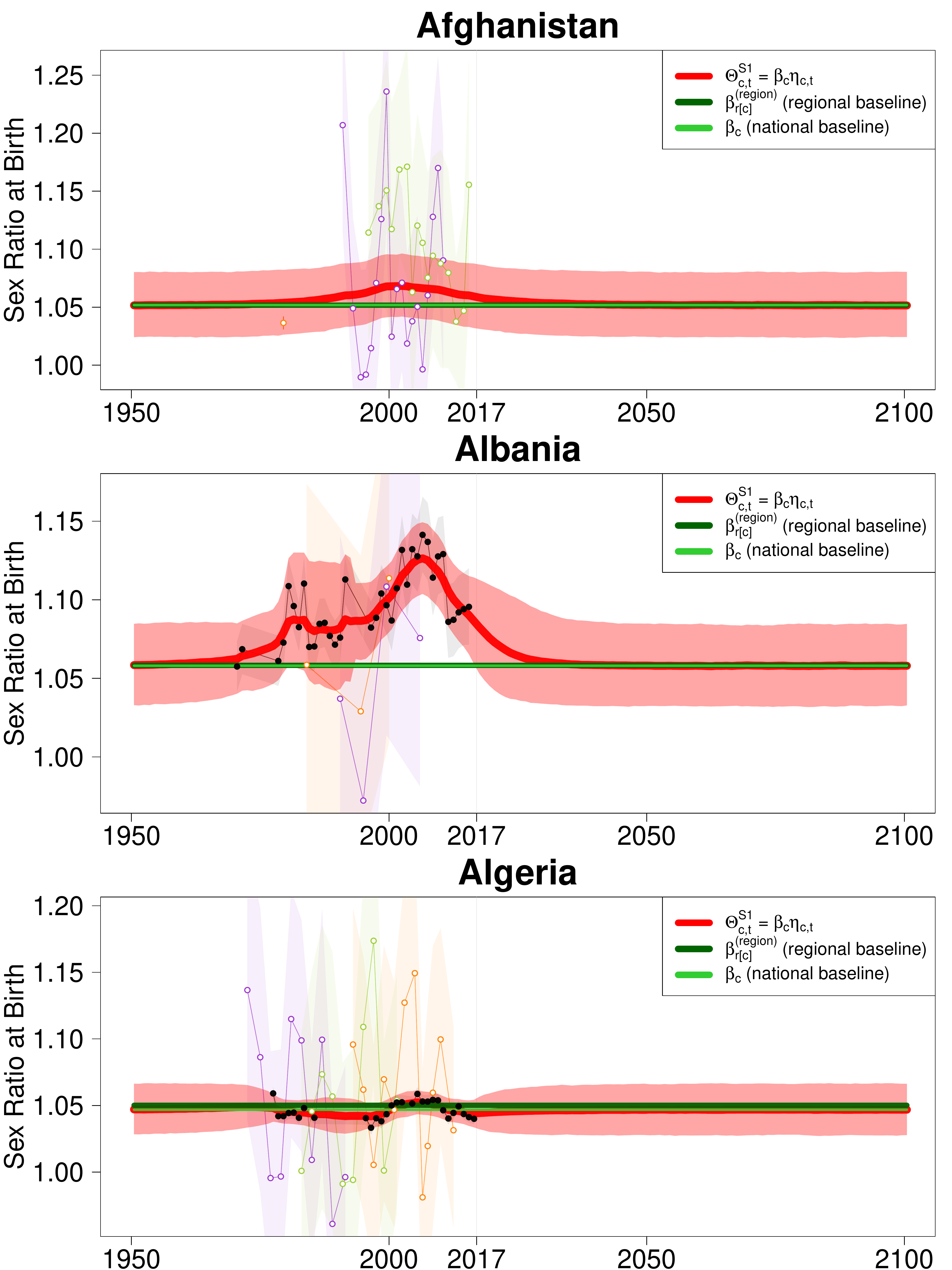}
}

\afterpage{
\pagenumbering{gobble}
\topskip0pt
\vspace*{\fill}
\centering
--- The End ---
\vspace*{\fill}
\newpage
}

\end{document}